\documentclass[sigconf]{acmart}
\AtBeginDocument{%
  }
\renewcommand\footnotetextcopyrightpermission[1]{} 
\usepackage{makecell}
\usepackage[ruled,vlined]{algorithm2e}
\usepackage{algpseudocode}
\usepackage{listings}
\usepackage{multirow}
\usepackage{subfigure}
\usepackage{tabularx}
\usepackage[table]{xcolor}
\usepackage{fontawesome}
\usepackage{siunitx}
\usepackage{bbding}
\usepackage{pifont}
\usepackage{xspace}
\usepackage{enumitem}
\setlist[itemize]{leftmargin=*}
\usepackage{amsmath}
\usepackage{booktabs}
\usepackage{float}
\usepackage{tikz}
\usetikzlibrary{arrows.meta,positioning,fit,backgrounds,shapes.geometric,calc,patterns,matrix}
\usepackage{pgfplots}
\pgfplotsset{compat=1.18}
\usepgfplotslibrary{groupplots}

\definecolor{cA}{HTML}{4C72B0}   
\definecolor{cB}{HTML}{C44E52}   
\definecolor{cC}{HTML}{55A868}   
\definecolor{cD}{HTML}{DD8452}   
\definecolor{cE}{HTML}{64B5CD}   
\definecolor{cF}{HTML}{8172B3}   
\definecolor{cG}{HTML}{A6A6A6}   
\definecolor{cInk}{HTML}{333333}
\definecolor{cGrid}{HTML}{E5E5E5}
\colorlet{lvSyn}{cA}   
\colorlet{lvSem}{cC}   
\colorlet{lvLog}{cD}   
\colorlet{cAgent}{cF}  
\colorlet{cOrc}{cG}    
\colorlet{fFT}{cA!70}      
\colorlet{fRL}{cC!70}      
\colorlet{fAgent}{cD!80}   
\colorlet{fLLM}{cG!70}     

\pgfplotsset{
  paperstyle/.style={
    font=\scriptsize,
    every tick label/.append style={font=\scriptsize, text=cInk},
    label style={font=\scriptsize, text=cInk},
    title style={font=\scriptsize\bfseries, text=cInk, yshift=-1mm},
    axis x line*=bottom, axis y line*=left,
    axis line style={cInk!55, line width=0.45pt},
    tick style={cInk!55, line width=0.4pt},
    major tick length=2pt,
    grid style={cGrid, line width=0.35pt},
    legend style={font=\scriptsize, draw=none, fill=none, text=cInk,
                  /tikz/every even column/.append style={column sep=6pt}},
    legend cell align=left,
    every node near coord/.append style={font=\tiny, text=cInk},
  },
  barlegend/.style={
    /pgfplots/legend image code/.code={\fill[##1, draw=none] (0pt,-2.2pt) rectangle (6pt,3.8pt);},
  },
  ours/.style={fill=cB, draw=cB!70!black, line width=0.3pt},
}

\usepackage[most]{tcolorbox}
\newtcolorbox{takeaway}{enhanced, frame hidden, colback=cA!7, arc=1.5pt,
  borderline west={2.2pt}{0pt}{cA!80!black}, left=6pt, right=5pt, top=2.5pt, bottom=2.5pt,
  fontupper=\small, before skip=4pt, after skip=2pt, before upper={\textbf{\textcolor{cA!80!black}{Takeaway.}}\ }}
\newtcolorbox[auto counter]{promptbox}[2][]{
  enhanced, breakable, colback=cA!3, colframe=cA!70!black, boxrule=0.5pt, arc=2pt,
  left=4pt, right=4pt, top=2pt, bottom=2pt, fontupper=\footnotesize,
  coltitle=white, colbacktitle=cA!80!black, fonttitle=\small\bfseries,
  title={Prompt~\thetcbcounter: #2}, before skip=3pt, after skip=3pt, #1}

\newcommand{\method}{SQLoom\xspace}
\newcommand{\dataset}{Spider-Refined\xspace}

\newcommand{\Rmnum}[1]{\uppercase\expandafter{\romannumeral #1}}

\newcommand{\mk}[1]{\textbf{\textcolor{cA!75!black}{#1}}}
\newcolumntype{L}[1]{>{\raggedright\arraybackslash}p{#1}}
\newcommand{\tabstyle}{\footnotesize\setlength{\aboverulesep}{0pt}\setlength{\belowrulesep}{0pt}\renewcommand{\arraystretch}{1.12}}
\newcommand{\headrow}{\rowcolor{cA!13}}
\newcommand{\grouprow}{\rowcolor{cG!14}}
\newcommand{\oursrow}{\rowcolor{cB!9}}
\newcommand{\best}[1]{\textbf{\textcolor{cB!85!black}{#1}}}
\newcommand{\second}[1]{\underline{#1}}
\newcommand{\yes}{\textcolor{cC!80!black}{\ding{51}}}
\newcommand{\no}{\textcolor{cG!80!black}{\ding{55}}}

\setcopyright{acmlicensed}
\copyrightyear{2027}
\acmYear{2027}
\acmDOI{XXXXXXX.XXXXXXX}
\acmConference[WWW '27]{the ACM Web Conference 2027}{May 10--14, 2027}{Dublin, Ireland}
\acmISBN{978-1-4503-XXXX-X/2027/XX}

\author{Hao Fu}
\email{haofu@vt.edu}
\affiliation{%
  \institution{Virginia Tech}
  \country{}
}

\author{Degang Sun}
\email{dgsun@cnic.cn}
\affiliation{%
  \institution{Computer Network Information Center, Chinese Academy of Sciences}
  \country{}
}

\author{Wan Wei}
\email{wanwei@cnic.cn}
\affiliation{%
  \institution{Computer Network Information Center, Chinese Academy of Sciences}
  \country{}
}

\author{Jinxia Wei}
\email{weijinxia@cnic.cn}
\affiliation{%
  \institution{Computer Network Information Center, Chinese Academy of Sciences}
  \country{}
}

\author{Chun Long}
\email{longchun@cnic.cn}
\affiliation{%
  \institution{Computer Network Information Center, Chinese Academy of Sciences}
  \country{}
}

\author{Peng Gao}
\email{penggao@vt.edu}
\affiliation{%
  \institution{Virginia Tech}
  \country{}
}

\begin{document}

\title{\method: Set-Contrastive Learning of Syntactic, Semantic, and Logical Dependencies for Text-to-SQL}

\begin{abstract}
Text-to-SQL translates natural language questions into SQL queries, enabling database access without writing SQL. Fine-tuning open-weight large language models (LLMs) is cost-effective, but existing methods treat SQL as ordinary text and optimize token-level likelihood without explicitly modeling its syntactic, semantic, and logical requirements. Thus, a single incorrect alias or comparison operator can invalidate a query while contributing little to training loss. We propose \method, a constraint-aware fine-tuning framework with supervision at three levels: syntactic validity under SQL grammar and database schema, semantic alignment between query decisions and the question, and logical consistency across dependent clause decisions. \method extends supervised fine-tuning with a unified set-contrastive objective favoring sets of correct outputs over minimally different incorrect ones at each level. LLM agents propose this supervision, which a SQL parser and execution on counterexample databases verify. We also release \dataset, a training dataset correcting 378 flawed Spider annotations and enriching database schemas. Experiments on Spider, BIRD, and three perturbed Spider variants show that \method improves execution accuracy for Qwen3-1.7B from 62.9\% zero-shot to 88.7\% on Spider development, reaching 89.2\% on Spider test and 73.4\% on BIRD development. It outperforms fine-tuning, reinforcement learning, and agent-based baselines with up to 70 billion parameters. On question perturbations, \method loses at most 0.6 accuracy points versus up to 12.1 for agent-based methods. It generates SQL in a single pass at 0.42\,s per query, 23$\times$ faster than a GPT-4-based agent method. These results demonstrate the effectiveness of set-contrastive supervision over token-level training.
\end{abstract}


\maketitle

\section{Introduction}
Text-to-SQL translates natural language questions into executable SQL queries~\cite{Yual18c,li2023can}, enabling users without SQL expertise to query relational databases through natural language interfaces. In recent years, large language models (LLMs) have been widely adopted in text-to-SQL research, owing to their strong natural language understanding~\cite{gao2024dail,hong2024survey}. However, general-purpose LLMs are not specifically trained for this task, and they often generate queries that reference non-existent columns, misinterpret the question, or combine inconsistent clauses. To mitigate these errors, prompt-based agent methods equip LLMs with extensive error-correction mechanisms, such as self-correction~\cite{pourreza2023din,askari2025magic}, multi-agent refinement~\cite{wang2025macsql,cen2025sqlfixagent,yuan2025cogsql}, and candidate selection~\cite{pourreza2024chase}. These mechanisms require multiple calls to proprietary LLMs per question, which increases latency and cost~\cite{zhang2025cost} and hinders deployment in Web services with high query volumes. Efficient fine-tuning methods that equip general-purpose LLMs with SQL-specific competence are therefore needed.

Existing fine-tuning methods for text-to-SQL fall into three main categories~\cite{hong2024survey}. \emph{Data-centric} methods enlarge the training corpus through SQL-centric pre-training~\cite{li2024codes} and data synthesis~\cite{yang2024sense,li2025omnisql}. Each sample still teaches a single reference string, so the model has to infer the rules of SQL from millions of samples, e.g., 2.5M for OmniSQL~\cite{li2025omnisql}. \emph{Task-decomposition} methods add auxiliary tasks or stages such as schema linking and skeleton parsing~\cite{li2023resdsql,pourreza2024dts,qin2025route,song2025jolt}. They help the model select tables and columns, but the query itself is still learned by imitating one reference token by token. \emph{Feedback-driven} methods optimize execution-based preferences or rewards with DPO or reinforcement learning~\cite{zhai2025excot,ma2025sqlr1,pourreza2025reasoningsql,yao2026arctic}. A reward judges the whole query, so it cannot tell which part caused an error, and it still rewards a wrong query when its error does not affect the result on the given database. Most of these methods treat SQL as ordinary text, and even structure-aware rewards~\cite{ma2026blindspots,lee2026expo} score a complete query instead of contrasting the tokens, decisions, and dependent clauses where errors occur.

\begin{figure}[t]
    \centering
    \includegraphics[width=\columnwidth]{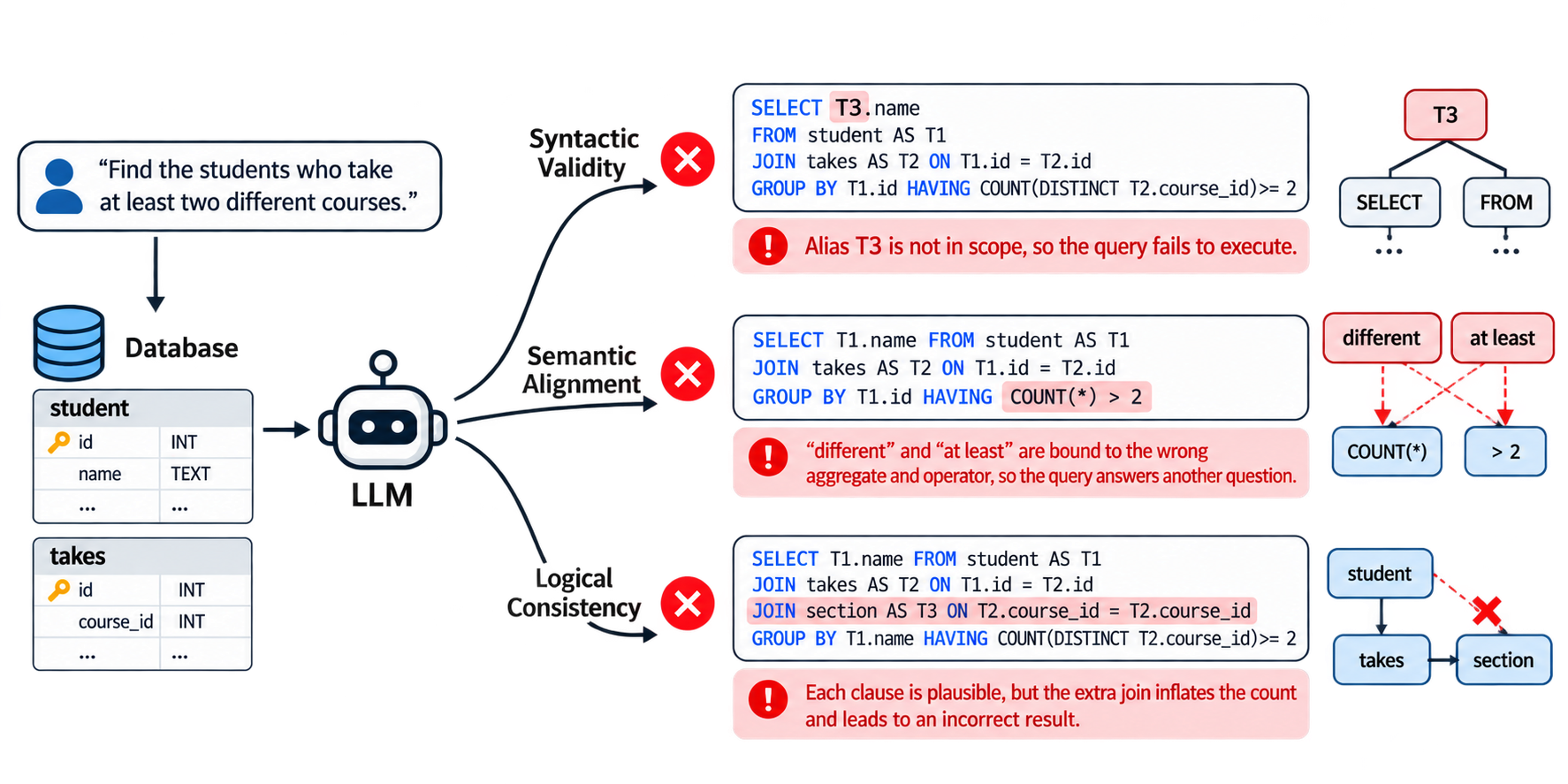}
    \caption{Three challenges of SQL generation on the running example. Each row shows a query with one type of error, and highlighted spans mark the wrong decisions.}
    \Description{A question with its reference SQL query, followed by three queries that violate syntactic validity, semantic alignment, and logical consistency.}
    \label{fig:intro}
\end{figure}

Without supervision from this structure, the current fine-tuning paradigm faces three challenges, illustrated in Figure~\ref{fig:intro}. \mk{C1} \emph{Syntactic validity.} A query must use only defined tables, columns, and aliases. The first query in Figure~\ref{fig:intro} selects \texttt{T3.name}, but no table is named \texttt{T3}. PICARD~\cite{scholak2021picard} filters such tokens with a parser during decoding, but the model still prefers them and needs the parser at deployment. \mk{C2} \emph{Semantic alignment.} Each phrase of the question must map to the correct SQL fragment. ``Different courses'' requires \texttt{COUNT(DISTINCT course\_id)} and ``at least two'' requires \texttt{>= 2}, whereas the second query writes \texttt{COUNT(*) > 2}. Schema-linking objectives~\cite{qin2025route,song2025jolt} teach the model which tables and columns to use, but not which aggregates and operators. \mk{C3} \emph{Logical consistency.} Dependent clauses must change together. The third query joins \texttt{section}, which yields one row per course section, but keeps \texttt{COUNT(*)}, so a course is counted several times. Each clause looks reasonable alone, and the query often returns the correct result on the test database, so execution rewards miss it.

To tackle these challenges, we propose \method, a constraint-aware fine-tuning framework. Our key insight is that SQL generation is better supervised by contrasting \emph{sets} of outputs than by imitating a single reference. Owing to the grammar and the schema, a question often has several correct queries, and at many positions the correct output is a class of tokens, e.g., the columns of the joined tables after \texttt{SELECT}, rather than one token. Token-level SFT rewards one member and pushes down the others, including correct ones. \method therefore augments SFT with a set-contrastive objective that favors a set of correct outputs over a set of wrong outputs differing from them at a single place, at three granularities. \textbf{Syntactic constraint for \mk{C1}.} At each position, the correct set contains the tokens allowed by the grammar and the schema, and the wrong set contains the rest. \textbf{Semantic constraint for \mk{C2}.} For each question phrase, the correct set contains its reference SQL fragment, e.g., \texttt{>= 2} for ``at least two'', and the wrong set contains fragments the model confuses with it, e.g., \texttt{> 2}. \textbf{Logical constraint for \mk{C3}.} For each pair of dependent clauses, we edit the question so that one clause must change, e.g., counting only courses with a lab section. The correct set contains the query that also updates the dependent clause, e.g., the count, and the wrong set contains the query that changes only the edited clause. The three constraints thus address the three challenges by teaching which tokens are legal, which fragments match the question, and which clauses must change together.

Our technical contributions are summarized below.
\begin{itemize}
    \item We propose \method, which replaces token-by-token imitation of one reference query with set-contrastive supervision labeled by a parser and query execution, instantiated as three constraints on syntax, semantics, and logic.
    \item We design a \emph{syntactic constraint} that uses the parse state as a training signal. An incremental parser with schema scope derives the legal tokens at each position, and a trie over schema identifiers aligns grammar terminals with subword tokens, so the model learns the grammar in training and needs no parser to correct its output at inference.
    \item We design a \emph{semantic constraint} that teaches the model to map each question phrase to the correct SQL fragment rather than a confusable one. It grounds question phrases in the AST nodes that realize them and contrasts each node with confusions mined from the model's own errors.
    \item We design a \emph{logical constraint} that keeps dependent clauses consistent when one of them changes. It derives these dependencies from a dependency graph over clauses and contrasts consistent and inconsistent queries on counterexample databases where their results differ.
\end{itemize}

Constructing these constraints requires correct and wrong sets for every training example, which are absent from the original annotations. The syntactic sets are derived directly by the parser, whereas the semantic and logical sets are built by a constraint synthesizer with four LLM agents. For the semantic constraint, an error-mining agent finds the model's wrong decisions and an alignment agent maps question phrases to SQL fragments with their confusions. For the logical constraint, an editing agent rewrites the question along a dependency edge with the two queries, and a verification agent checks the edited query's intent. Each output must pass the parser and execution checks. While synthesizing constraints in our experiments, the error-mining agent also flagged 378 Spider annotations that mislabel correct predictions or teach incorrect queries. After manual review, we correct these annotations and release \dataset, a training dataset that extends Spider with the corrected annotations, schema descriptions generated by the synthesizer, and the verified constraint samples. \dataset serves only as training data, and all evaluations use the official labels. Experiments lead to four findings.
\begin{itemize}
    \item \textbf{The gains come from the objective.} With the same training data, replacing the constraints with plain SFT lowers EX by 8.8 points on Spider Test and 10.3 points on BIRD Dev.
    \item \textbf{A 1.7B model outperforms much larger systems.} \method reaches 88.7\% EX on Spider Dev (62.9\% zero-shot), 89.2\% on Spider Test, and 73.4\% on BIRD Dev, above all evaluated baselines and zero-shot Qwen3-235B and GPT-4.
    \item \textbf{\method is robust to perturbed questions.} It loses at most 0.6 points on the three Spider variants, whereas agent-based methods lose 8.8--12.1 points on Spider-DK.
    \item \textbf{\method is efficient.} It generates SQL in one pass (0.42\,s per query), 23$\times$ faster than CogSQL with GPT-4 in our setup.
\end{itemize}

Overall, these results show that with constraint-based supervision, even a small LLM can generate SQL accurately and robustly.

\section{Background}
\label{sec:related}

\subsection{Problem Formulation}
\label{sec:formulation}
Given a natural language question $q$ and a database $\mathcal{D}$ with schema $\mathcal{S}$ (tables, columns, primary and foreign keys), text-to-SQL generates a query $y=(y_1,\dots,y_T)$ whose execution result on $\mathcal{D}$ answers $q$. Let $x=(q,\mathcal{S})$ denote the model input. Fine-tuning methods train an LLM $p_\theta$ by minimizing the negative log-likelihood of a reference query $y^{*}$,
\begin{equation}
    \mathcal{L}_{\mathrm{SFT}}=-\sum_{t=1}^{T}\log p_\theta\left(y^{*}_t \mid x, y^{*}_{<t}\right).
    \label{eq:sft}
\end{equation}

\subsection{Related Work}
\noindent\textbf{Prompt-based Agent Methods.}
Early neural parsers encode the schema with task-specific architectures~\cite{zhong2017seq2sql,wang2020ratsql,li2023graphix}. LLM-based methods instead improve a general model at inference, through example selection, decomposition, and schema pruning~\cite{gao2024dail,pourreza2023din,xie2024decomposition,qu2024before,talaei2024chess,li2024dawn}, or through agents that plan, refine, and select among candidates~\cite{wang2025macsql,yuan2025cogsql,askari2025magic,cen2025sqlfixagent,pourreza2024chase,shkapenyuk2025automatic}. Other work checks SQL with tools rather than with a model. Constrained decoding runs an incremental SQL parser during generation and discards tokens that violate the grammar or the schema~\cite{scholak2021picard}, and test-suite evaluation executes a predicted query on several generated databases to expose queries that are correct only by coincidence on the original one~\cite{zhong2020testsuite}. In essence, these methods repair or detect errors after generation instead of preventing them. The repair is judged by the same LLM or by execution failures, so a query that executes but answers another question often passes, and each question pays for several LLM calls~\cite{zhang2025cost}. The tools also act outside training, so the model does not learn from them and the parser must stay in the serving loop.

\noindent\textbf{Data-centric Fine-tuning.}
This direction scales the training corpus through SQL-centric pre-training and data synthesis~\cite{sun2023sql,li2024codes,yang2024sense,li2025omnisql}. Its limitation lies in the form of supervision rather than its amount (Table~\ref{tab:compare}). Each sample still provides one reference string, so the model must infer grammar rules, question-to-SQL bindings, and clause dependencies statistically, and equivalent queries for one question even provide conflicting targets.

\noindent\textbf{Task-decomposition Fine-tuning.}
This direction splits generation into schema linking, skeleton parsing, and SQL stages, or adds them as auxiliary tasks~\cite{li2023resdsql,pourreza2024dts,qin2025route,song2025jolt}. It strengthens the recognition of the input, namely which tables and columns to use (Table~\ref{tab:compare}), but leaves output-side decisions, such as aggregates, operators, and their consistency across clauses, to token-level imitation.

\noindent\textbf{Feedback-driven Fine-tuning.}
This direction optimizes execution-based preferences or rewards with DPO or reinforcement learning~\cite{rafailov2023dpo,zhai2025excot,liu2025cotdpo,ma2025sqlr1,pourreza2025reasoningsql,yao2026arctic}, and recent work refines the reward with graph, AST, or clause-level matching~\cite{weng2025graphreward,ma2026blindspots,lee2026expo}. The core problem is the signal itself. An outcome-level reward spreads the same credit over all tokens, and single-database execution wrongly rewards inconsistent queries with correct results (Table~\ref{tab:compare}).

\begin{table}[t]
  \centering
  \tabstyle
  \caption{Supervision signals of representative fine-tuning methods. Syn., Sem., and Log. denote explicit supervision of syntactic validity, semantic alignment beyond tables and columns, and cross-clause consistency.}
  \label{tab:compare}
  \setlength{\tabcolsep}{3pt}
  \resizebox{\columnwidth}{!}{%
  \begin{tabular}{@{}llccc@{}}
    \toprule
    \headrow \textbf{Method} & \textbf{Training signal} & \textbf{Syn.} & \textbf{Sem.} & \textbf{Log.} \\
    \midrule
    CodeS~\cite{li2024codes}, OmniSQL~\cite{li2025omnisql} & token likelihood & \no & \no & \no \\
    ROUTE~\cite{qin2025route}, JOLT-SQL~\cite{song2025jolt} & + schema linking & \no & \no & \no \\
    ExCoT~\cite{zhai2025excot} & query-level preference & \no & \no & \no \\
    SQL-R1~\cite{ma2025sqlr1}, Arctic~\cite{yao2026arctic} & query-level reward & \no & \no & \no \\
    Ma et al.~\cite{ma2026blindspots} & AST-matching reward & \yes & \no & \no \\
    EXPO-SQL~\cite{lee2026expo} & clause-level reward & \yes & \no & \no \\
    \oursrow \textbf{\method} & unified set-contrastive objective & \yes & \yes & \yes \\
    \bottomrule
  \end{tabular}}
\end{table}

\subsection{Motivation}
\label{sec:motivation}
\noindent\textbf{Problem measurement.} To measure which errors remain under standard fine-tuning, we fine-tune Qwen3-1.7B on the Spider training set with LoRA and token-level SFT, and classify its 211 model errors on the Spider development set (Figure~\ref{fig:errors}; categories in Appendix~\ref{app:errors}). Wrong columns dominate from medium to extra-hard questions, followed by semantic errors. These queries are valid but bind a question element to a wrong decision, which typically spans only one or a few tokens.

\begin{figure}[t]
    \centering
\begin{tikzpicture}
\begin{groupplot}[
    group style={group size=2 by 1, horizontal sep=14mm},
    paperstyle,
    scale only axis, height=2.1cm,
    y dir=reverse, y axis line style={draw=none}, ytick style={draw=none},
    xmajorgrids,
]
\nextgroupplot[title={(a) Errors per difficulty level}, width=2.5cm,
    xbar stacked, /pgf/bar width=8.5pt,
    xmin=0, xmax=112, xtick={0,25,50,75,100}, xlabel={Number of errors}, clip=false,
    symbolic y coords={Easy,Medium,Hard,Extra},
    ytick={Easy,Medium,Hard,Extra},
    yticklabels={Easy\,(470),Medium\,(857),Hard\,(463),Extra\,(357)},
    enlarge y limits=0.17,
    legend to name=errleg, legend columns=3, barlegend,
    legend style={/tikz/every even column/.append style={column sep=5pt}}]
\addplot[fill=cE!90, draw=white, line width=0.3pt] coordinates {(3,Easy) (16,Medium) (6,Hard) (12,Extra)};
\addplot[fill=cA!85, draw=white, line width=0.3pt] coordinates {(0,Easy) (36,Medium) (38,Hard) (54,Extra)};
\addplot[fill=cC!80, draw=white, line width=0.3pt] coordinates {(4,Easy) (8,Medium) (13,Hard) (18,Extra)};
\addplot[fill=cD!85, draw=white, line width=0.3pt] coordinates {(0,Easy) (1,Medium) (0,Hard) (1,Extra)};
\addplot[fill=cF!80, draw=white, line width=0.3pt] coordinates {(0,Easy) (0,Medium) (1,Hard) (0,Extra)};
\legend{Wrong table, Wrong column, Semantic error, Wrong ordering or \texttt{LIMIT}, Other model error}
\node[font=\scriptsize, text=cInk, anchor=west] at (axis cs:7,Easy) {7};
\node[font=\scriptsize, text=cInk, anchor=west] at (axis cs:61,Medium) {61};
\node[font=\scriptsize, text=cInk, anchor=west] at (axis cs:58,Hard) {58};
\node[font=\scriptsize, text=cInk, anchor=west] at (axis cs:85,Extra) {85};
\nextgroupplot[title={(b) Share by error type (\%)}, width=1.8cm,
    xbar, /pgf/bar width=6pt, /pgf/bar shift=0pt,
    xmin=0, xmax=78, xtick={0,20,40,60},
    symbolic y coords={C,S,T,Q,O},
    ytick={C,S,T,Q,O},
    yticklabels={Wrong column, Semantic, Wrong table, Ordering/\texttt{LIMIT}, Other},
    enlarge y limits=0.12,
    nodes near coords, every node near coord/.append style={font=\scriptsize, anchor=west, xshift=-1pt},
    point meta=explicit symbolic]
\addplot[fill=cA!85, draw=none] coordinates {(60.7,C) [60.7]};
\addplot[fill=cC!80, draw=none] coordinates {(20.4,S) [20.4]};
\addplot[fill=cE!90, draw=none] coordinates {(17.5,T) [17.5]};
\addplot[fill=cD!85, draw=none] coordinates {(0.9,Q) [0.9]};
\addplot[fill=cF!80, draw=none] coordinates {(0.5,O) [0.5]};
\end{groupplot}
\node[anchor=south] at ($(group c1r1.north west)!0.5!(group c2r1.north east)+(0,5.5mm)$) {\pgfplotslegendfromname{errleg}};
\end{tikzpicture}
    \caption{Model errors of LoRA fine-tuning on Spider Dev (a) per difficulty level, with question counts, and (b) by type.}
    \Description{Left, horizontal stacked bars of error counts by type for four difficulty levels. Right, horizontal bars of the share of each error type.}
    \label{fig:errors}
\end{figure}
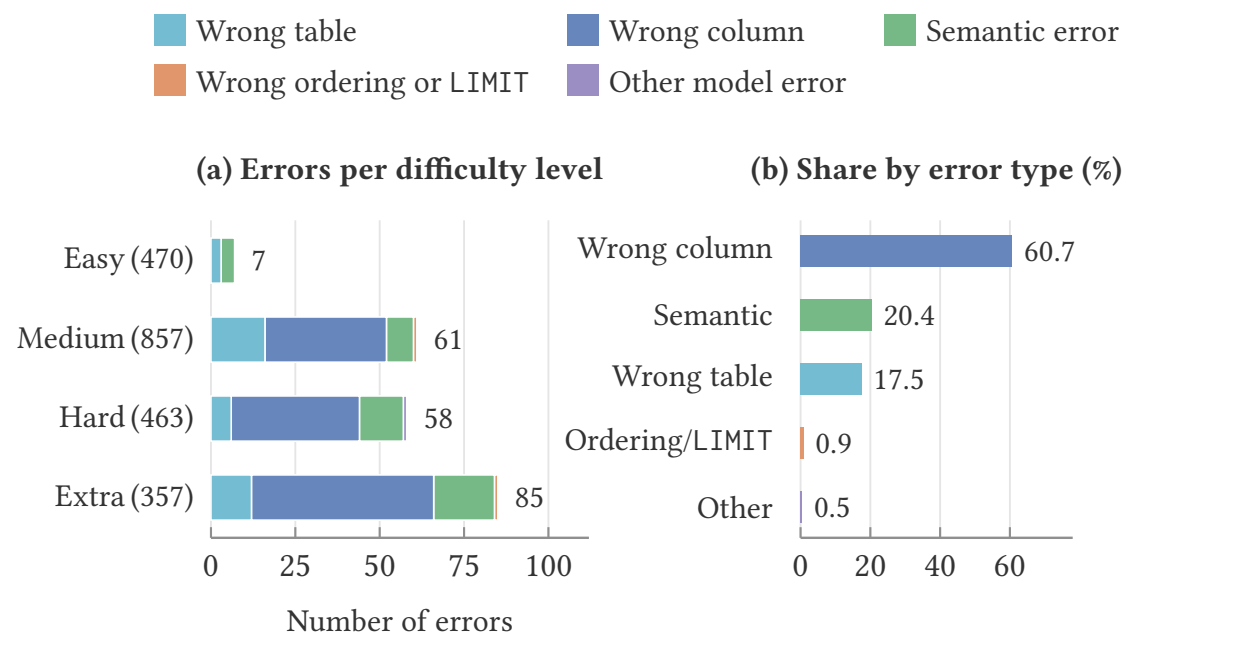

These model errors trace back to Eq.~\ref{eq:sft}, which raises the likelihood of the reference without explicitly penalizing incorrect outputs and ignores three properties of SQL. 1) \emph{Validity is decidable.} Given the grammar and the schema, an incremental parser determines which tokens are legal at each position, yet SFT does not explicitly lower the probability of illegal ones. 2) \emph{Errors are local but decisive.} A wrong decision, such as \texttt{COUNT(*)} instead of \texttt{COUNT(DISTINCT\,$\cdot$)}, occupies a small share of the SFT loss but changes the answer; conversely, since SQL labels are not unique, an equivalent rewriting of $y^{*}$ is penalized as much as a wrong query. 3) \emph{Decisions are dependent.} A one-to-many join may require \texttt{DISTINCT}, and \texttt{GROUP BY} must follow the target entity, but a query that violates such a dependency can still return the correct result on the given database.

\noindent\textbf{Requirements.} A fine-tuning objective for SQL should therefore meet three requirements. First, it contrasts the reference with outputs \emph{certified} to be wrong by a parser or by execution beyond the given database. Second, it confines the contrast to the structural unit in which the outputs differ. Third, it applies at the level of tokens, decisions, and dependent decisions.

\section{Methodology}
\label{sec:method}

\subsection{Overview}
\label{sec:unified}
\begin{figure*}[t]
    \centering
    \includegraphics[width=\textwidth]{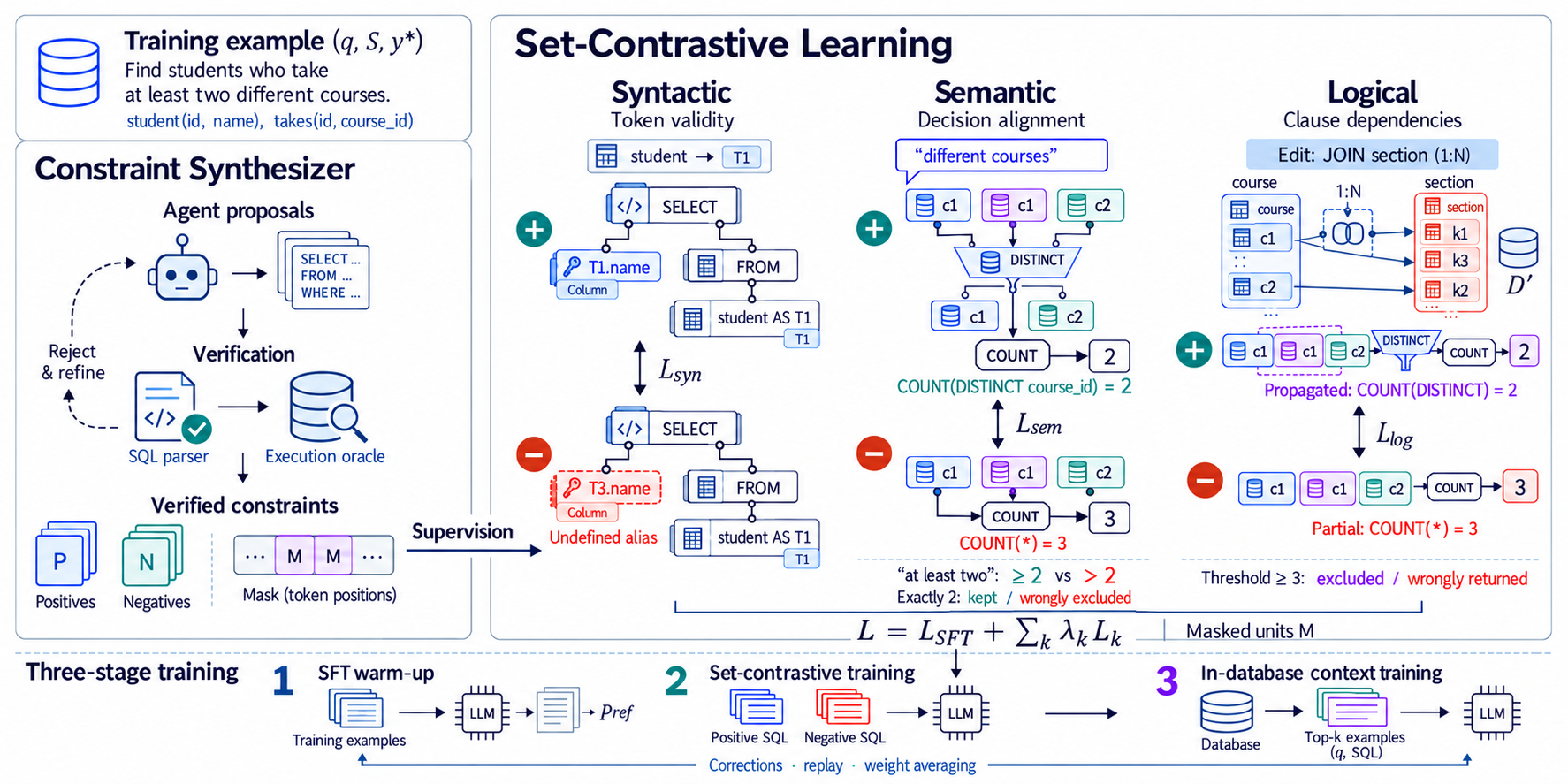}
    \caption{Overview of \method. The constraint synthesizer builds positive and negative sets, a set-contrastive loss favors positives (colored) over minimally different negatives (red) at three levels, and training proceeds in three stages.}
    \Description{A question and a database schema feed an LLM, whose outputs are contrasted against positive and negative sets at the syntactic, semantic, and logical levels.}
    \label{fig:2}
\end{figure*}

Token-level SFT maximizes the likelihood of the reference without penalizing incorrect outputs, so wrong tokens, wrong decisions, and inconsistent clauses receive little training signal. \method supervises these errors explicitly while keeping single-pass inference. As shown in Figure~\ref{fig:2}, the constraint synthesizer (Section~\ref{sec:engine}) builds sets of correct outputs and minimally different wrong outputs at the syntactic, semantic, and logical levels. A set-contrastive objective favors the correct set over the wrong set at each of the three levels. A three-stage procedure trains the model with this objective.

\noindent\textbf{Three-level correctness hierarchy.} The three challenges occur at different granularities, so one loss on the whole query cannot tell where an error lies. We therefore define one unit per challenge, namely a \emph{token} for \mk{C1}, a \emph{decision} for \mk{C2}, and a \emph{pair of dependent decisions} for \mk{C3}. A decision is the SQL fragment that realizes one element of the question, e.g., \texttt{>= 2} for ``at least two'', and corresponds to a node of the abstract syntax tree (AST) of the query. A logical unit links two decisions in different clauses, e.g., a join and the count that depends on it.

\noindent\textbf{Oracle-certified constraints.} Since one question can have several correct queries, comparing an output with the reference cannot tell whether the output is wrong or merely written differently. For each unit $u$ we therefore build a constraint $c=(\mathcal{P}_c,\mathcal{N}_c,M_c)$, where $\mathcal{P}_c$ is a set of correct (\emph{admissible}) outputs, $\mathcal{N}_c$ is a set of wrong (\emph{inadmissible}) outputs that differ from them only in $u$, and $M_c(y)$ marks the token positions of $y$ that belong to $u$. Wrong outputs are certified by an \emph{oracle}, i.e., a program with a definite answer rather than a learned model, whereas correct outputs are references or pass additional checks. The oracle is a schema-aware SQL parser at the syntactic level. At the other two levels it executes the queries on the original database $\mathcal{D}$ and on counterexample databases $\{\mathcal{D}'_j\}$, i.e., small generated databases on which similar queries return different results. Appendix~\ref{app:impl} details both oracles.

\noindent\textbf{Set-contrastive objective.} We compare outputs only at the positions of their unit. Let $r_\theta(y,M)=\sum_{t\in M}\log\frac{p_\theta(y_t\mid x,y_{<t})}{p_{\mathrm{ref}}(y_t\mid x,y_{<t})}$ be how much more likely the model makes $y$ at positions $M$ than a frozen reference model $p_{\mathrm{ref}}$, here the model after warm-up. Each constraint is enforced by
\begin{equation}
    \ell(c)=-\log\frac{\sum_{y\in\mathcal{P}_c}\exp\big(\beta\, r_\theta(y,M_c(y))\big)}{\sum_{y\in\mathcal{P}_c\cup\mathcal{N}_c}\exp\big(\beta\, r_\theta(y,M_c(y))\big)},
    \label{eq:unified}
\end{equation}
which raises the total probability of the correct set over the wrong set at the positions of the unit, while the rest of the query keeps the plain SFT signal. The temperature $\beta$ controls how sharply the two sets are separated. The full objective is
\begin{equation}
    \mathcal{L}=\mathcal{L}_{\mathrm{SFT}}+\sum_{k\in\{\mathrm{syn},\mathrm{sem},\mathrm{log}\}}\lambda_{k}\,\mathcal{L}_{k},\qquad \mathcal{L}_{k}=\sum_{c\in\mathcal{C}_{k}}\ell(c),
    \label{eq:overall}
\end{equation}
where $\mathcal{C}_k$ is the set of constraints at level $k$, and $\mathcal{L}_{\mathrm{SFT}}$ covers the references and all admissible outputs. Table~\ref{tab:levels} lists the instantiations.

\begin{table}[t]
  \centering
  \tabstyle
  \caption{Instantiations of the set-contrastive constraint (Eq.~\ref{eq:unified}) at the three levels. The last row lists each level's negatives.}
  \label{tab:levels}
  \setlength{\tabcolsep}{2.5pt}
  \begin{tabular}{@{}>{\raggedright\arraybackslash}p{1.15cm}>{\raggedright\arraybackslash}p{1.85cm}>{\raggedright\arraybackslash}p{2.3cm}>{\raggedright\arraybackslash}p{2.4cm}@{}}
    \toprule
    \headrow & \textbf{Syntactic} & \textbf{Semantic} & \textbf{Logical} \\
    \midrule
    Unit $u$ & decision position $t$ & AST node $d_i$ aligned to element $e_i$ & dependent decision pair (edge of $G$) \\
    $\mathcal{P}_c$ & valid next tokens & reference decision & propagated $y'$ \\
    $\mathcal{N}_c$ & invalid next tokens & verified confusions & partially propagated $\tilde{y}'$ \\
    $M_c$ & position $t$ & tokens of $d_i$ & tokens of affected clauses \\
    Oracle & parser & execution on $\mathcal{D},\mathcal{D}'$ & execution on $\mathcal{D},\mathcal{D}'$ \\
    \grouprow Negatives & invalid & valid, misaligned & valid, aligned, inconsistent \\
    \bottomrule
  \end{tabular}
\end{table}

Semantic negatives are valid queries, and logical negatives also keep each decision consistent with the question (last row of Table~\ref{tab:levels}), so each level covers errors that the previous levels miss. The loss never penalizes equivalent outputs and updates only the constrained unit (Appendix~\ref{app:props}).

\subsection{Syntactic Constraint}
\label{sec:syntax}
The syntactic constraint teaches the model which tokens may come next (\mk{C1}). Legality depends on the grammar and the schema, e.g., an alias defined inside a subquery is not visible outside it, and SQL keywords and names do not align with the LLM's subword tokens.

We feed $y^{*}$ token by token into an incremental parser, i.e., a SQL parser that can analyze an unfinished query, extended with the schema. After each prefix $y^{*}_{<t}$, the parser returns a state $s_t$ that records what may come next and which tables, aliases, and columns are visible. The legal set $\mathcal{V}_{\mathrm{valid}}(s_t)$ contains each token $v$ after which the query can still be completed into a valid one, e.g., only table names of $\mathcal{S}$ after \texttt{FROM}. Since a name may span several subword tokens, legality is checked on the decoded text with a prefix tree (trie) of keywords and schema names, so the partial name \texttt{stu} of \texttt{student} counts as legal. The constraint is applied at decision positions $\mathcal{T}_{d}$, i.e., the first token of each keyword, name, operator, or parenthesis, and the legal sets are cached per database.

The syntactic instance of Eq.~\ref{eq:unified} is
\begin{equation}
    \mathcal{L}_{\mathrm{syn}}=-\sum_{t\in\mathcal{T}_{d}}\log\sum_{v\in\mathcal{V}_{\mathrm{valid}}(s_t)}p_\theta\left(v\mid x,y^{*}_{<t}\right),
    \label{eq:syntax}
\end{equation}
which moves probability from illegal tokens to the whole legal set rather than to the reference token alone, so legal alternatives are not pushed down. The model thus learns validity during training and needs no parser at inference.

\subsection{Semantic Constraint}
\label{sec:align}
The semantic constraint binds each element of the question to the SQL decision that realizes it (\mk{C2}). Datasets do not annotate which phrase maps to which part of the query, and a seemingly wrong alternative may be an equivalent rewrite.

For each training example, we extract alignments $\mathcal{A}=\{(e_i,d_i)\}$, where $e_i$ is an element of $q$, e.g., ``at least two'', and $d_i$ is the AST node of $y^{*}$ that realizes it, e.g., \texttt{>= 2} (Figure~\ref{fig:align}). Elements fall into six types, namely entity, attribute, value, aggregation, comparison/quantifier, and ordering. An alignment is kept only if replacing $d_i$ by another option of the same type changes the execution result, which confirms that $d_i$ depends on $e_i$. As an auxiliary task, the model writes the alignment list before the SQL, supervised by $\mathcal{L}_{\mathrm{SFT}}$.

For each alignment, $\mathcal{P}_c=\{y^{*}\}$, and $\mathcal{N}_c$ contains confusions $\tilde{y}$ that replace $d_i$ with a similar but wrong option, e.g., a similar column, \texttt{COUNT(*)} for \texttt{COUNT(DISTINCT\,$\cdot$)}, or \texttt{>} for \texttt{>=}. Candidates are ranked by how often the current model makes them, so training targets its actual mistakes. Only valid confusions that return a different result from $y^{*}$ are kept, so equivalent rewrites are not penalized by design. With one confusion per unit, Eq.~\ref{eq:unified} reduces to the pairwise loss of direct preference optimization (DPO)~\cite{rafailov2023dpo} restricted to the tokens of $d_i$, which corrects the wrong decision without changing the rest of the query.

\subsection{Logical Constraint}
\label{sec:dep}
The logical constraint teaches the model to update dependent clauses together (\mk{C3}). A single training pair rarely shows how one clause should change when another changes, and an inconsistent query often returns the correct result on $\mathcal{D}$, so executing it there does not reveal the error.

\begin{figure}[t]
    \centering
    \includegraphics[width=\columnwidth]{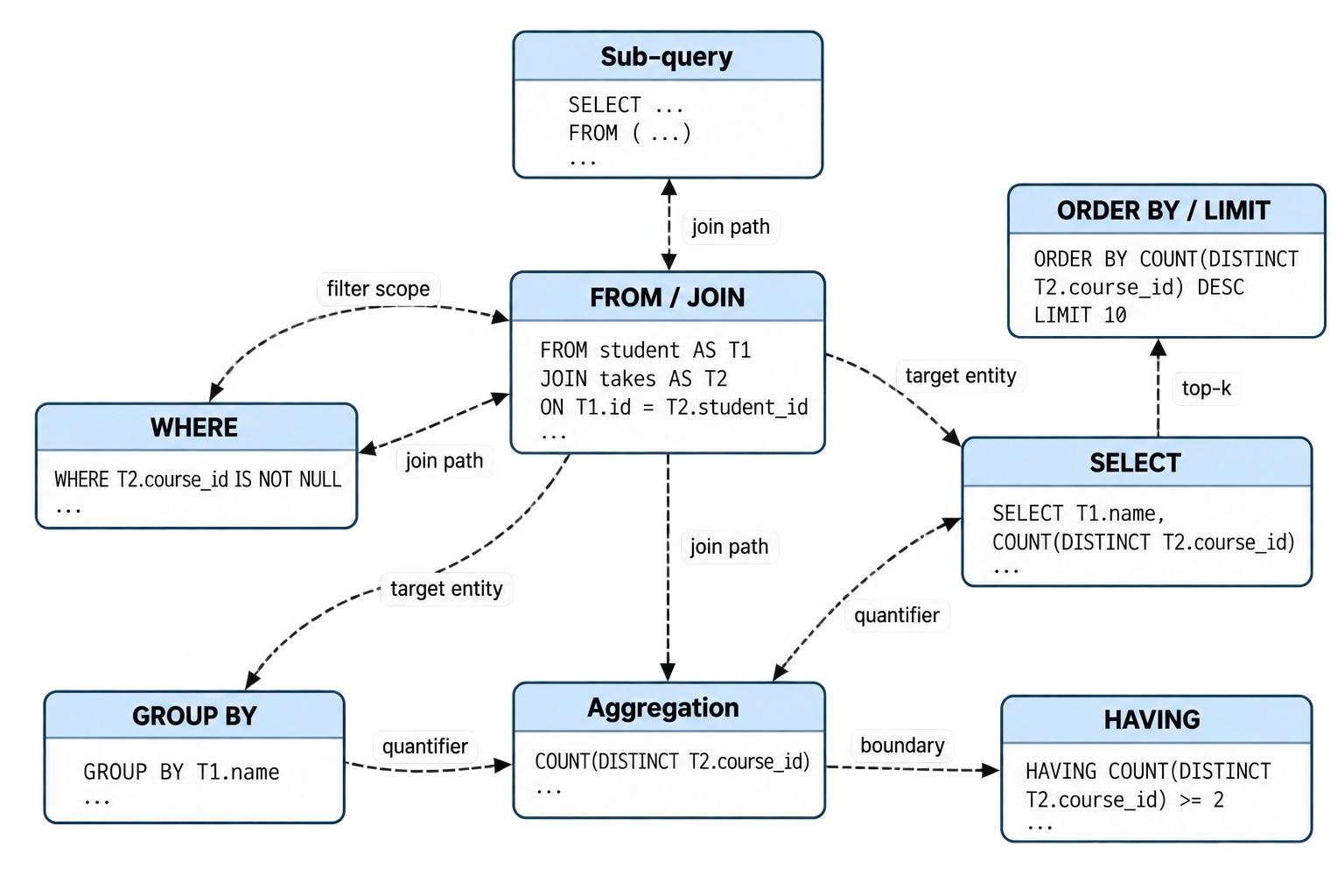}
    \caption{Component dependency graph $G$ on an example query. Each labeled edge marks a component that must be checked when the source component changes.}
    \Description{Graph of SQL components connected by labeled dependency edges.}
    \label{fig:dep}
\end{figure}

We split a query into components (\texttt{FROM/JOIN}, \texttt{WHERE}, \texttt{GROUP BY}, aggregation, \texttt{HAVING}, \texttt{SELECT}, \texttt{ORDER BY/LIMIT}, and sub-queries) and build a graph $G$ in which an edge states that a certain change to one component requires checking another (Figure~\ref{fig:dep}; all edges in Table~\ref{tab:dep}). For example, adding a one-to-many join requires checking whether the count still counts each entity once. Each edge is a unit. Edges mark what may change, and execution decides how.

Starting from $(q,y^{*})$, we change one aligned element of the question and generate an edited question $q'$, a propagated query $y'$ in which the edited component and the components that depend on it are updated, and a partially propagated query $\tilde{y}'$ in which only the edited component changes. Then $\mathcal{P}_c=\{y'\}$ and $\mathcal{N}_c=\{\tilde{y}'\}$ for input $(q',\mathcal{S})$. For example, restricting the running example to courses with a lab section adds a join with \texttt{section}; $y'$ counts \texttt{DISTINCT T2.course\_id}, whereas $\tilde{y}'$ keeps \texttt{COUNT(*)}, which counts a course per lab section. The pair isolates the dependency and is kept only if the two queries differ on a counterexample database, $y'$ runs on all databases, and the verification agent judges $y'$ consistent with $q'$.

The mask $M_c$ covers the components that differ between $y'$ and $\tilde{y}'$. Since both queries contain the edit, the loss focuses on the components that must change \emph{because of} it. Eq.~\ref{eq:unified} again reduces to a DPO-style loss on these positions, and $y'$ is also added to $\mathcal{L}_{\mathrm{SFT}}$. Appendices~\ref{app:method}, \ref{app:impl}, and~\ref{app:examples} show examples of all levels.

\subsection{Training Procedure}
\label{sec:schedule}
\label{sec:context}
The training procedure combines the three constraints in one model. Preference losses are unstable before the model can write reasonable SQL, and later fine-tuning stages tend to overwrite earlier skills. \method therefore trains in three stages (Algorithm~\ref{alg:train}). Stage~1 is a short SFT warm-up whose result also serves as $p_{\mathrm{ref}}$. Stage~2 trains with Eq.~\ref{eq:overall}. Stage~3 prepends to each input the $k$ most similar verified question--query pairs of the same database, since Web applications answer many questions over one database. After each stage, the model is further fine-tuned on queries corrected by the error-mining agent $A_e$ (Section~\ref{sec:engine}), mixed with a share of earlier examples (replay), and its weights are averaged with the pre-trained ones as in WiSE-FT~\cite{wortsman2022robust} to reduce forgetting.

\begin{algorithm}[t]
  \caption{Three-stage training procedure of \method.}
  \label{alg:train}
  \small
  \SetKwFunction{Correct}{Correct}
  \SetKwProg{Fn}{Function}{:}{}
  \KwIn{training set $\mathcal{T}$, pre-trained LLM $M_p$}
  $M\leftarrow$ \Correct{SFT of $M_p$ on $\mathcal{T}$ for one epoch} \tcp*{Stage 1}
  $p_{\mathrm{ref}}\leftarrow M$; build $\mathcal{C}_{\mathrm{syn}},\mathcal{C}_{\mathrm{sem}},\mathcal{C}_{\mathrm{log}}$ with the constraint synthesizer\;
  \ForEach{mini-batch $\mathcal{B}$ of $\mathcal{T}\cup\mathcal{C}_{\mathrm{syn}}\cup\mathcal{C}_{\mathrm{sem}}\cup\mathcal{C}_{\mathrm{log}}$}{
    update $M$ by minimizing $\mathcal{L}$ (Eq.~\ref{eq:overall}) on $\mathcal{B}$ \tcp*{Stage 2}
  }
  $M\leftarrow$\Correct{$M$}\;
  $M\leftarrow$ \Correct{SFT with top-$k$ in-database examples} \tcp*{Stage 3}
  \Return{$M$}\;
  \Fn{\Correct{$M$}}{
    train $M$ with $\mathcal{L}_{\mathrm{SFT}}$ on verified corrections of $A_e(M,\mathcal{T})$ and a replay of $p_{\mathrm{sample}}$ earlier examples\;
    merge LoRA adapters; \Return{$\gamma M+(1-\gamma)M_p$}\;
  }
\end{algorithm}

\section{Verified Constraint Synthesis}
\label{sec:dataengine}
\subsection{Agent-based Constraint Synthesizer}
\label{sec:engine}
\begin{figure}[t]
    \centering
    \includegraphics[width=0.9\columnwidth]{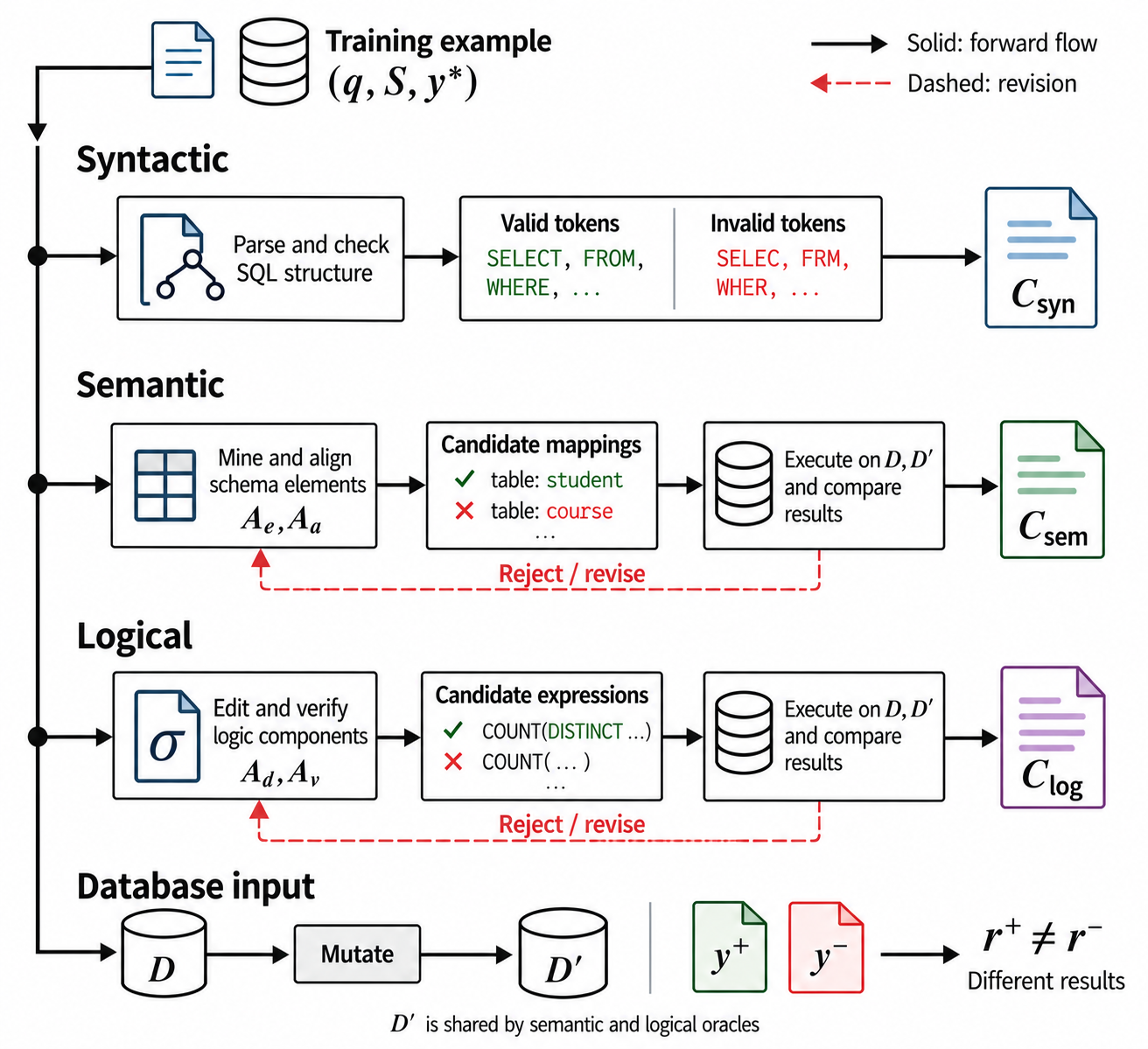}
    \caption{Agent-based constraint synthesizer. Agents or a parser propose candidates in three lanes, LLM-free oracles accept or reject them (dashed), and counterexample databases $\mathcal{D}'$ serve the semantic and logical lanes.}
    \Description{Three lanes, each with an agent, candidate constraints, and an oracle, feed accepted syntactic, semantic, and logical constraints. A counterexample generator on the left supplies databases to the semantic and logical lanes.}
    \label{fig:engine}
\end{figure}

The constraint synthesizer instantiates the constraints of Section~\ref{sec:unified}. As shown in Figure~\ref{fig:engine}, four LLM agents propose candidates, and the grammar and execution oracles decide their membership in $\mathcal{P}_c$ or $\mathcal{N}_c$. The parser alone yields the syntactic constraint, $A_e$ and $A_a$ the semantic one, and $A_d$ and $A_v$ the logical one.

\begin{itemize}
    \item \textbf{Error-mining agent $A_e$.} After each stage, $A_e$ finds the wrong decision in each failed training prediction and writes a close correction, repaired by execution for at most $i^{c}_{\max}$ rounds. Corrections serve as SFT examples, the wrong decisions select the units for semantic and logical constraints, and suspicious references are flagged for manual review.
    \item \textbf{Alignment agent $A_a$} proposes semantic units (alignments) and up to three confusable alternatives per decision.
    \item \textbf{Editing agent $A_d$} samples a logical unit (an edge of $G$) and produces $q'$, $y'$, and $\tilde{y}'$.
    \item \textbf{Verification agent $A_v$} back-translates $y'$ into a question and checks whether it matches $q'$.
\end{itemize}
\noindent\textbf{Counterexample database generation.} Given a database $\mathcal{D}$ and two queries $y_1$ and $y_2$ with the same result on $\mathcal{D}$, the generator searches for a database $\mathcal{D}'$ that satisfies the constraints of $\mathcal{D}$ and on which $y_1$ and $y_2$ return different results. It (1) samples a sub-database containing the tables referenced by the queries, following foreign keys to keep referential integrity; (2) mutates it by duplicating rows with fresh primary keys, perturbing values at the boundaries of query constants (e.g., $2$ in \texttt{>= 2}), and inserting additional child rows into one-to-many relationships; and (3) keeps a mutated instance if it satisfies the primary-key, foreign-key, \texttt{NOT NULL}, and type constraints and separates the two queries. Duplicated child rows expose join-induced double counting, boundary values expose inclusive versus exclusive comparisons, and ties expose superlative and \texttt{LIMIT} errors. The search stops when a separating instance is found or a mutation budget is exhausted. The same procedure is used for the non-equivalence check of confusions and the substitution test of alignments. Each training example contributes at most one pair per edge type in Table~\ref{tab:dep}.

Because each candidate must pass checks that do not depend on an LLM, agent errors reduce the amount of data rather than its correctness. For example, an incorrect alignment fails the substitution test, an incorrect propagated query is not separated from its partial counterpart, and an equivalent confusion is discarded. The agent prompts are listed in Appendix~\ref{app:prompts}.

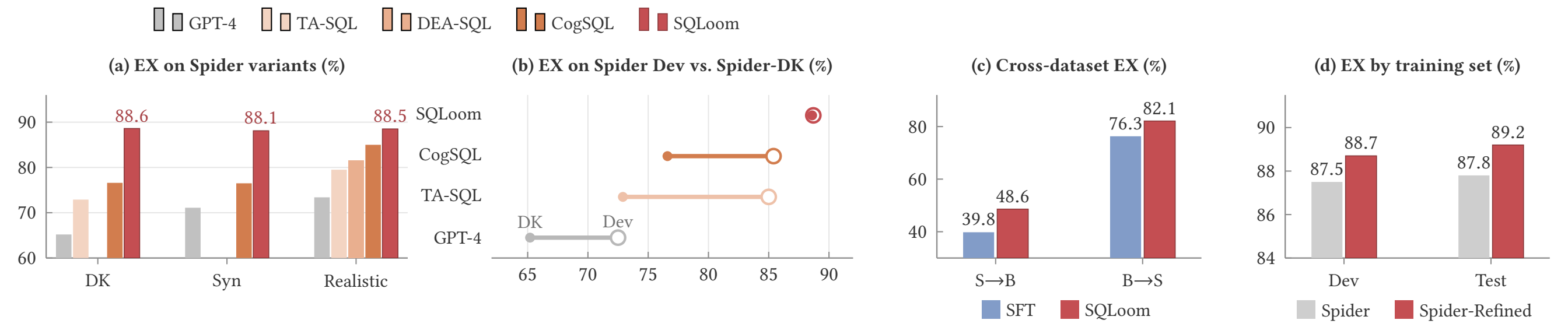
\begin{figure*}[t]
    \centering
    \begin{tikzpicture}
\begin{groupplot}[
    group style={group size=4 by 1, horizontal sep=9.5mm},
    paperstyle,
    scale only axis, height=1.85cm,
    barlegend,
]
\nextgroupplot[title={(a) EX on Spider variants (\%)}, width=4.1cm, ybar=0.5pt, /pgf/bar width=5pt,
    symbolic x coords={DK,Syn,Realistic}, xtick={DK,Syn,Realistic}, ymajorgrids,
    ymin=60, ymax=96, ytick={60,70,80,90}, enlarge x limits=0.2,
    legend to name=robleg, legend columns=5]
\addplot[fill=fLLM, draw=none] coordinates {(DK,65.2) (Syn,71.1) (Realistic,73.4)};
\addplot[fill=cD!35, draw=none] coordinates {(DK,72.9) (Realistic,79.5)};
\addplot[fill=cD!65, draw=none] coordinates {(Realistic,81.5)};
\addplot[fill=cD!95!black, draw=none] coordinates {(DK,76.6) (Syn,76.5) (Realistic,85.0)};
\addplot[ours, nodes near coords, every node near coord/.append style={font=\scriptsize, text=cB!85!black, anchor=south, yshift=-1pt}] coordinates {(DK,88.6) (Syn,88.1) (Realistic,88.5)};
\legend{GPT-4, TA-SQL, DEA-SQL, CogSQL, \method}
\nextgroupplot[title={(b) EX on Spider Dev vs.\ Spider-DK (\%)}, width=4.1cm,
    xmin=62, xmax=92, xtick={65,70,75,80,85,90}, xmajorgrids,
    ymin=0.5, ymax=4.5, ytick={1,2,3,4}, yticklabels={\method, CogSQL, TA-SQL, GPT-4},
    y dir=reverse,
    y axis line style={draw=none}, ytick style={draw=none}]
\addplot[draw=none, forget plot] coordinates {(62,1)};
\foreach \y/\dk/\dev/\col in {4/65.2/72.5/cG!80, 3/72.9/85.0/cD!50, 2/76.6/85.4/cD!95!black, 1/88.6/88.7/cB} {
  \edef\temp{\noexpand\draw[\col, line width=1.4pt] (axis cs:\dk,\y) -- (axis cs:\dev,\y);
    \noexpand\draw[\col, fill=white, line width=0.8pt] (axis cs:\dev,\y) circle (2.3pt);
    \noexpand\fill[\col] (axis cs:\dk,\y) circle (1.6pt);}
  \temp
}
\node[font=\scriptsize, text=cInk!70, anchor=south] at (axis cs:72.5,4) {Dev};
\node[font=\scriptsize, text=cInk!70, anchor=south] at (axis cs:65.2,4) {DK};
\nextgroupplot[title={(c) Cross-dataset EX (\%)}, width=3.0cm, /pgf/bar width=10pt,
    symbolic x coords={s2b,b2s}, xtick={s2b,b2s}, xticklabels={S$\rightarrow$B, B$\rightarrow$S},
    ymin=30, ymax=92, ytick={40,60,80}, enlarge x limits=0.4, ybar=1pt, barlegend,
    nodes near coords, every node near coord/.append style={font=\scriptsize, anchor=south, yshift=-1pt},
    legend style={at={(0.5,-0.2)}, anchor=north, legend columns=2, font=\scriptsize}]
\addplot[fill=fFT, draw=none] coordinates {(s2b,39.8) (b2s,76.3)};
\addplot[ours] coordinates {(s2b,48.6) (b2s,82.1)};
\legend{SFT, \method}
\nextgroupplot[title={(d) EX by training set (\%)}, width=3.0cm, /pgf/bar width=10pt,
    symbolic x coords={Dev,Test}, xtick={Dev,Test}, xticklabels={Dev, Test},
    ymin=84, ymax=91.5, ytick={84,86,88,90}, enlarge x limits=0.4, ybar=1pt, barlegend,
    nodes near coords, every node near coord/.append style={font=\scriptsize, anchor=south, yshift=-1pt},
    legend style={at={(0.5,-0.2)}, anchor=north, legend columns=2, font=\scriptsize}]
\addplot[fill=cG!55, draw=none] coordinates {(Dev,87.5) (Test,87.8)};
\addplot[ours] coordinates {(Dev,88.7) (Test,89.2)};
\legend{Spider, \dataset}
\end{groupplot}
\node[anchor=south] at ($(group c1r1.north)!0.5!(group c2r1.north)+(0,4.5mm)$) {\pgfplotslegendfromname{robleg}};
\end{tikzpicture}
    \caption{Robustness and transfer. (a) EX on the Spider variants. (b) EX on Spider Dev (hollow) and Spider-DK (filled); a shorter segment means higher robustness. (c) Cross-dataset EX (S = Spider, B = BIRD). (d) EX when training on Spider vs.\ \dataset.}
    \Description{Four bar charts of accuracy on Spider variants, accuracy drop, cross-dataset transfer, and training data comparison.}
    \label{fig:robust}
\end{figure*}

\subsection{The \dataset Dataset}
\label{sec:data}
During error mining, many apparent model errors turned out to be Spider annotation problems. We manually reviewed all cases flagged by $A_e$ and introduce \dataset, a training dataset built from Spider that corrects these annotations, enriches the database schemas, and adds the verified alignments, confusions, and counterfactual pairs produced by the constraint synthesizer. \dataset is used only for training; evaluation uses the official sets.

Among the confirmed annotation issues, the most frequent is a correct prediction that is rejected by execution-based comparison (163 cases), followed by under-specified questions whose references return extra (71) or missing (37) columns, and references whose logic contradicts the question (31). Figure~\ref{fig:rv} lists all issue types. For each confirmed issue, the reference query or the question is minimally revised; when several outputs are equally valid, the alternatives are recorded. All revisions are verified by execution.

Spider schemas often use terse or ambiguous identifiers. For each database, the synthesizer agents generate descriptions of (i) table semantics, e.g., \texttt{player} stores player records; (ii) value types, e.g., \texttt{REAL} for continuous values such as income and \texttt{INTEGER} for counts; (iii) relationships, e.g., one club has many players; (iv) potential pitfalls, e.g., a floating-point primary key that makes equality comparison unreliable; and (v) intended use cases, e.g., retrieving the players of a club or aggregating income per club. The descriptions are checked manually and added to the schema input $\mathcal{S}$ and the agent prompts.

\section{Experiments}
\label{sec:exp}
We evaluate how \method compares with fine-tuning, reinforcement learning, and agent-based methods and with general-purpose LLMs, its robustness to perturbed questions and unseen datasets, how each component contributes, and its cost.

\subsection{Experimental Settings}
\textbf{Datasets.} We use \textbf{Spider}~\cite{Yual18c} (10,181 questions, 200 databases, 138 domains), its variants \textbf{Spider-DK}~\cite{gan2021exploring}, \textbf{Spider-Syn}~\cite{gan2021towards}, and \textbf{Spider-Realistic}~\cite{deng2020structure}, and \textbf{BIRD}~\cite{li2023can} (12,751 pairs, 95 databases, 33.4\,GB). We train two models, one on the \dataset training set, evaluated on Spider and its variants, and one on the BIRD training set, evaluated on BIRD. Evaluation databases are disjoint from training databases, so all evaluations are cross-database. Spider difficulty levels follow the official script, which counts SQL components such as joins, aggregations, and nested queries.

\noindent\textbf{Baselines.} We compare with fine-tuning methods (T5-3B+PICARD~\cite{scholak2021picard}, RESDSQL~\cite{li2023resdsql}, SQL-PaLM~\cite{sun2023sql}, CodeS~\cite{li2024codes}, SENSE~\cite{yang2024sense}, ROUTE~\cite{qin2025route}, OmniSQL~\cite{li2025omnisql}, and JOLT-SQL~\cite{song2025jolt}), preference and RL methods (ExCoT~\cite{zhai2025excot}, SQL-R1~\cite{ma2025sqlr1}, and Arctic-Text2SQL-R1~\cite{yao2026arctic}), and agent-based methods (SQLFixAgent~\cite{cen2025sqlfixagent}, DEA-SQL~\cite{xie2024decomposition}, SuperSQL~\cite{li2024dawn}, CogSQL~\cite{yuan2025cogsql}, TA-SQL~\cite{qu2024before}, and AskData~\cite{shkapenyuk2025automatic}).

\noindent\textbf{Metrics and implementation.} We report execution accuracy (EX). \method fine-tunes Qwen3-1.7B with LoRA~\cite{hu2022lora} on one RTX 4090 GPU and decodes greedily without constrained decoding. The agents use GPT-4. We set $\lambda_{\mathrm{syn}}=0.5$, $\lambda_{\mathrm{sem}}=\lambda_{\mathrm{log}}=1.0$, $\beta=0.1$, and $k=3$, selected by grid search on a held-out 10\% of the training set (Appendix~\ref{app:impl}). Results of \method are averaged over 3 seeds.

\subsection{Main Results}
\label{sec:main}
\begin{table}[t]
    \centering
    \tabstyle
    \caption{EX (\%) on Spider and BIRD. The best results are in bold red and the second best are underlined, and $^{\dagger}$ denotes self-consistency. \method uses no BIRD evidence and reports mean\,$\pm$\,std over three seeds.}
    \label{tab:res_dedicated}
    \setlength{\tabcolsep}{9.5pt}
    \begin{tabular}{@{}lccc@{}}
        \toprule
        \headrow & \multicolumn{2}{c}{\textbf{Spider}} & \textbf{BIRD} \\
        \headrow \multirow{-2}{*}{\textbf{Method}} & \textbf{Dev} & \textbf{Test} & \textbf{Dev} \\ \midrule
        \grouprow \multicolumn{4}{@{}l}{\emph{Fine-tuning}} \\
        T5-3B + PICARD & 79.3 & 75.1 & -- \\
        RESDSQL-3B + NatSQL & 84.1 & 79.9 & -- \\
        SQL-PaLM & 82.8 & -- & -- \\
        CodeS-7B & 85.4 & 83.5 & 57.2 \\
        SENSE-7B & 83.2 & 83.5 & 51.8 \\
        ROUTE + Qwen2.5-7B & 83.6 & 83.7 & 55.9 \\
        OmniSQL-7B & 81.2 & 87.9 & 63.9 \\
        JOLT-SQL + Qwen2.5-Coder-7B & 87.0 & 86.8 & 60.4 \\
        \grouprow \multicolumn{4}{@{}l}{\emph{Preference / reinforcement learning}} \\
        ExCoT + LLaMA-3.1-70B & -- & 86.6 & 68.5 \\
        SQL-R1 + Qwen2.5-Coder-7B$^{\dagger}$ & \second{87.6} & 88.7 & 66.6 \\
        Arctic-Text2SQL-R1-7B & -- & \second{88.8} & \second{68.9} \\
        \grouprow \multicolumn{4}{@{}l}{\emph{Agent-based}} \\
        CodeS-7B + SQLFixAgent & 86.2 & 83.9 & -- \\
        DEA-SQL + GPT-4 & 85.4 & 87.1 & 58.9 \\
        SuperSQL + GPT-4 & 87.0 & -- & 58.5 \\
        CogSQL + GPT-4 & 85.4 & 86.4 & 59.6 \\
        TA-SQL + GPT-4 & 85.0 & -- & 56.2 \\
        AskData + GPT-4o & -- & -- & 67.4 \\ \midrule
        \oursrow \textbf{\method} + Qwen3-1.7B & \best{88.7}\scriptsize{$\pm$0.3} & \best{89.2}\scriptsize{$\pm$0.2} & \best{73.4}\scriptsize{$\pm$0.5} \\
        \bottomrule
    \end{tabular}
\end{table}

\begin{figure*}[t]
    \centering
    \begin{tikzpicture}
\begin{groupplot}[
    group style={group size=4 by 1, horizontal sep=9.5mm},
    paperstyle,
    scale only axis, height=2.3cm,
]
\nextgroupplot[title={(a) EX drop of ablations (pts)}, width=3.8cm,
    xbar=0.4pt, /pgf/bar width=3.4pt,
    xmin=0, xmax=40, xtick={0,10,20,30,40}, xmajorgrids,
    symbolic y coords={ctx,cons,corr,sft,nosyn,nosem,nolog,sftonly,mix},
    ytick={ctx,cons,corr,sft,nosyn,nosem,nolog,sftonly,mix},
    yticklabels={w/o context aug., w/o constraint training, w/o error-mined corr., w/o SFT warm-up,
      w/o $\mathcal{L}_{\mathrm{syn}}$, w/o $\mathcal{L}_{\mathrm{sem}}$,
      w/o $\mathcal{L}_{\mathrm{log}}$, Stage 2 = SFT only,
      single-stage mix},
    y dir=reverse, enlarge y limits=0.06,
    y axis line style={draw=none}, ytick style={draw=none},
    barlegend, legend style={at={(0.98,0.03)}, anchor=south east, legend columns=1, font=\scriptsize, fill=white, draw=cGrid, inner sep=1pt, row sep=-1.5pt}]
\addplot[fill=cA!80, draw=none] coordinates {(1.2,ctx) (12.5,cons) (11.1,corr) (27.4,sft) (0.9,nosyn) (2.3,nosem) (2.1,nolog) (8.8,sftonly) (3.6,mix)};
\addplot[fill=cA!35, draw=none] coordinates {(3.9,ctx) (9.7,cons) (17.2,corr) (33.6,sft) (1.5,nosyn) (3.2,nosem) (3.6,nolog) (10.3,sftonly) (5.3,mix)};
\legend{Spider Test, BIRD Dev}
\draw[cInk!30, densely dashed] ([yshift=-2.6mm]axis cs:0,sft) -- ([yshift=-2.6mm]axis cs:40,sft);
\nextgroupplot[title={(b) Spider Test EX by difficulty (\%)}, width=3.6cm,
    ymin=10, ymax=100, ytick={20,40,60,80,100}, ymajorgrids,
    xmin=-0.25, xmax=3.25, xtick={0,1,2,3}, xticklabels={Base, +S1, +S2, +S3},
    every axis plot/.append style={line width=0.8pt, mark size=1.4pt},
    legend style={at={(0.98,0.03)}, anchor=south east, legend columns=1, font=\scriptsize, fill=white, draw=cGrid, inner sep=1pt, row sep=-1.5pt}]
\addplot[cA, mark=*] coordinates {(0,72.0) (1,73.9) (2,91.2) (3,93.3)};
\addplot[cC!85!black, mark=square*] coordinates {(0,56.0) (1,57.2) (2,88.8) (3,89.6)};
\addplot[cD, mark=triangle*] coordinates {(0,32.0) (1,47.0) (2,87.2) (3,88.4)};
\addplot[cF, mark=diamond*] coordinates {(0,18.3) (1,40.0) (2,83.0) (3,83.7)};
\addplot[cG!90!black, densely dashed, mark=none] coordinates {(0,48.3) (1,55.8) (2,88.0) (3,89.2)};
\legend{Easy, Medium, Hard, Extra, All}

\nextgroupplot[title={(c) Training time (GPU hours)}, width=2.2cm, ybar stacked, /pgf/bar width=11pt,
    ymin=0, ymax=13, ytick={0,4,8,12}, ymajorgrids,
    symbolic x coords={SFT,Ours}, xtick={SFT,Ours}, xticklabels={SFT,\method}, enlarge x limits=0.55,
    barlegend, legend style={at={(0.02,0.98)}, anchor=north west, legend columns=1, font=\scriptsize, fill=white, draw=cGrid, inner sep=1pt, row sep=-1.5pt}]
\addplot[fill=fFT, draw=white] coordinates {(SFT,4.6) (Ours,1.2)};
\addplot[fill=cB!85, draw=white] coordinates {(SFT,0) (Ours,6.5)};
\addplot[fill=cB!40, draw=white] coordinates {(SFT,0) (Ours,3.1)};
\legend{S1 / SFT, S2, S3}
\node[font=\scriptsize, anchor=south] at (axis cs:SFT,4.6) {4.6};
\node[font=\scriptsize, anchor=south] at (axis cs:Ours,10.8) {10.8};

\nextgroupplot[title={(d) Inference latency (s/query)}, width=2.6cm, xbar, /pgf/bar width=6pt, /pgf/bar shift=0pt,
    xmode=log, xmin=0.1, xmax=40, xtick={0.1,1,10}, xticklabels={0.1,1,10}, log origin=infty,
    symbolic y coords={ours,sft,sqlr1,cog}, ytick={ours,sft,sqlr1,cog},
    yticklabels={\method, SFT, SQL-R1, CogSQL},
    y dir=reverse, enlarge y limits=0.2, xmajorgrids, y axis line style={draw=none}, ytick style={draw=none},
    point meta=explicit symbolic, nodes near coords, every node near coord/.append style={font=\scriptsize, anchor=west}]
\addplot[fill=fFT, draw=none] coordinates {(0.40,sft) [0.40]};
\addplot[fill=fRL, draw=none] coordinates {(2.9,sqlr1) [2.9]};
\addplot[fill=fAgent, draw=none] coordinates {(9.8,cog) [9.8]};
\addplot[ours] coordinates {(0.42,ours) [0.42]};
\end{groupplot}
\end{tikzpicture}
    \caption{Ablation and cost. (a) EX drop of ablated stages (above the dashed line) and constraint levels (below). (b) Spider Test EX by difficulty after Stages 1 to 3 (S1 to S3). (c) Training time. (d) Per-query latency (log scale).}
    \Description{A horizontal bar chart of ablation drops, a line chart of accuracy by difficulty across stages, and two bar charts of training time and inference latency.}
    \label{fig:analysis}
\end{figure*}
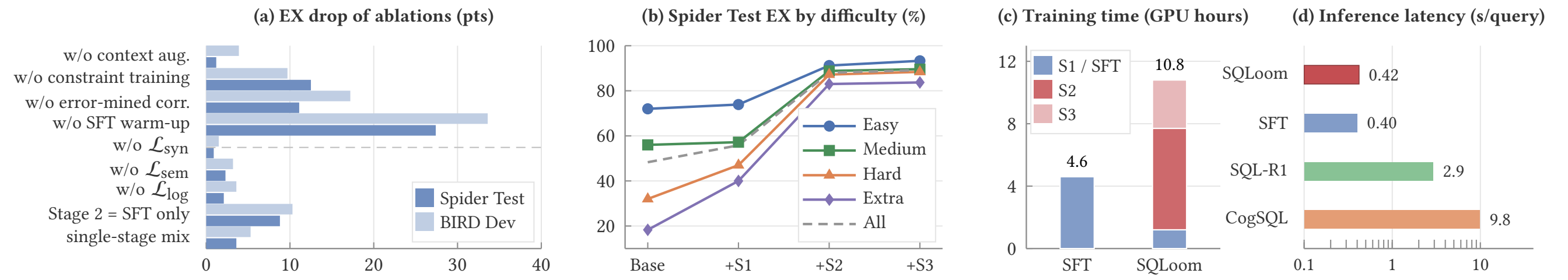

As shown in Table~\ref{tab:res_dedicated}, \method achieves the best EX on all three sets with the smallest model, exceeding the strongest baselines by 1.1, 0.4, and 4.5 points. It outperforms JOLT-SQL, the best fine-tuning method, by 13.0 points on BIRD, and AskData, the best agent-based method, by 6.0 points, without proprietary LLM calls.

As Table~\ref{tab:res_dedicated} and Figure~\ref{fig:sizeacc} show, all baselines above 85\% on the Spider test set or 60\% on BIRD use at least 7B parameters, whereas \method uses 1.7B. Its margin is largest on BIRD, whose questions involve more clauses and more chances for inconsistency.

Unlike most baselines, \method does not use the external knowledge evidence of BIRD; adding it raises EX to 76.8\%. The gains are also not explained by imitating the teacher, since oracles vet agent outputs and GPT-4 is weaker on the Spider variants (Figure~\ref{fig:robust}a).

\begin{takeaway}Explicit SQL supervision can matter more than scale: a small model trained to reject near misses beats much larger ones.\end{takeaway}

\subsection{Comparison with General-Purpose LLMs}
\label{sec:llms}
\begin{figure}[t]
    \centering
\begin{tikzpicture}
\begin{axis}[
    paperstyle,
    xbar, /pgf/bar width=3.8pt, /pgf/bar shift=0pt,
    scale only axis, width=5.0cm, height=3.05cm,
    xmin=0, xmax=100, xtick={0,25,50,75,100},
    xlabel={EX on Spider Dev (\%)},
    symbolic y coords={ours,q235,q8,gpt4,q17,dsv3,g4,q06,spark,coder,r18,r1671,seed,gem2,g27,g1},
    ytick={ours,q235,q8,gpt4,q17,dsv3,g4,q06,spark,coder,r18,r1671,seed,gem2,g27,g1},
    yticklabels={\textbf{\textcolor{cB!85!black}{Qwen3-1.7B + \method}},Qwen3-235B,Qwen3-8B,GPT-4,\textbf{Qwen3-1.7B},DeepSeek-V3,Gemma3-4B,Qwen3-0.6B,Spark-X1,Qwen2.5-Coder-7B,DeepSeek-R1-8B,DeepSeek-R1-671B,Seed-Coder-8B-R,Gemini-2.0,Gemma3-27B,Gemma3-1B},
    y dir=reverse, enlarge y limits=0.035,
    xmajorgrids, y axis line style={draw=none}, ytick style={draw=none},
    nodes near coords, every node near coord/.append style={font=\scriptsize, anchor=west, xshift=-1pt},
    legend style={at={(0.98,0.03)}, anchor=south east, fill=white, draw=cGrid, font=\scriptsize},
    barlegend,
]
\addplot[fill=cA!55, draw=none] coordinates {(75.3,q235) (75.0,q8) (60.8,dsv3) (42.0,g4) (40.3,q06) (9.6,r18) (2.9,r1671) (0,g27) (0,g1)};
\addplot[fill=fLLM, draw=none] coordinates {(72.5,gpt4) (35.2,spark) (0,gem2)};
\addplot[fill=cF!65, draw=none] coordinates {(22.6,coder) (0.1,seed)};
\addplot[ours] coordinates {(88.7,ours)};
\addplot[fill=cA!90!black, draw=none, forget plot] coordinates {(62.9,q17)};
\legend{Open-weight, Proprietary, Code-specialized, Fine-tuned}
\draw[->, cB!80!black, line width=0.6pt, densely dashed] ([xshift=6mm]axis cs:62.9,q17) to[out=0, in=0, looseness=0.9] node[pos=0.5, right, font=\scriptsize, text=cB!85!black] {+25.8} ([xshift=6mm]axis cs:88.7,ours);
\end{axis}
\end{tikzpicture}
    \caption{Zero-shot EX of general-purpose LLMs on Spider Dev, compared with Qwen3-1.7B fine-tuned by \method (top bar).}
    \Description{Horizontal bar chart of zero-shot accuracy of open-weight, proprietary, and code-specialized LLMs, with the fine-tuned model on top.}
    \label{fig:3}
\end{figure}
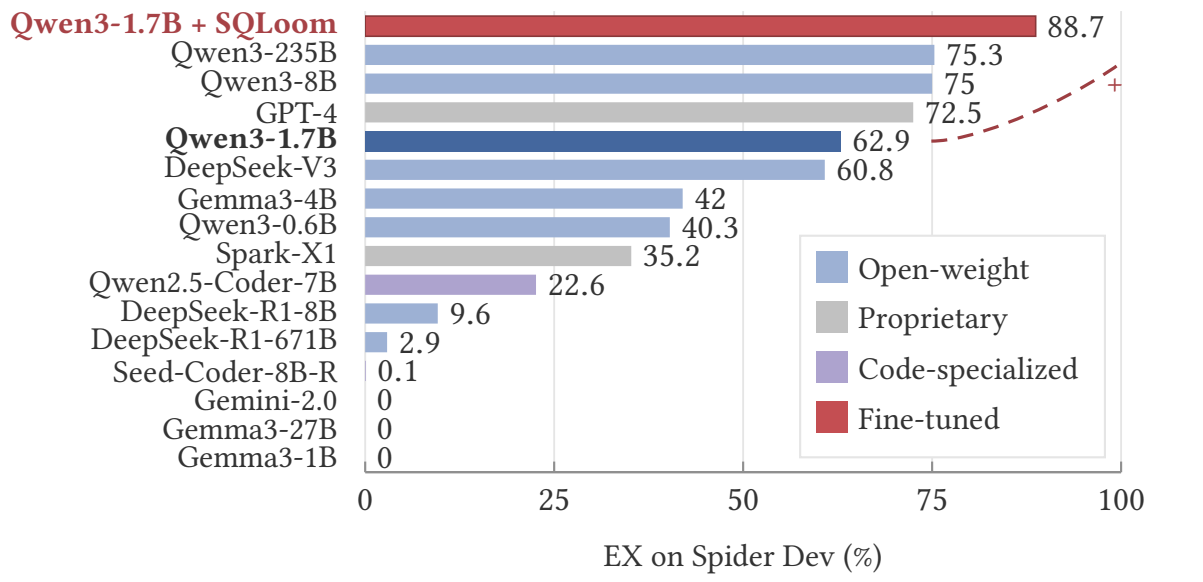
Figure~\ref{fig:3} shows that model size does not determine zero-shot accuracy. Qwen3-8B matches Qwen3-235B, and the open-weight DeepSeek-V3 is 14.4 points lower. Reasoning and code-specialized models score lowest; most of their failures lack a final SQL statement within the generation budget, which is why the longer-reasoning DeepSeek-R1-671B scores below DeepSeek-R1-8B. \method lifts Qwen3-1.7B from 62.9\% to 88.7\%, i.e., 13.4 points above the best zero-shot model, which is 138$\times$ larger.

\subsection{Robustness and Transfer}
\label{sec:robust}
On Spider-DK, Spider-Syn, and Spider-Realistic (Figure~\ref{fig:robust}a,b), \method loses at most 0.6 points relative to Spider Dev, whereas agent-based methods lose 8.8--12.1 points on Spider-DK and 8.9 points on Spider-Syn. Lexical matching between question and schema breaks under domain knowledge and synonyms, whereas \method grounds question elements in verified decisions.

In the cross-dataset setting (Figure~\ref{fig:robust}c), \method improves over SFT trained on the same data by 8.8 points from Spider to BIRD and 5.8 points from BIRD to Spider. Training on \dataset instead of Spider improves EX on the \emph{official} sets by 1.2 and 1.4 points (Figure~\ref{fig:robust}d), so the corrections help beyond evaluation.

\begin{takeaway}Supervising properties of SQL itself, rather than patterns of one benchmark, carries over to new phrasings and datasets.\end{takeaway}

\subsection{Ablation Study}
\label{sec:ablation}
All training stages contribute. Removing constraint training costs 12.5 and 9.7 points, and removing error-mined correction costs 11.1 and 17.2 points (Figure~\ref{fig:analysis}a). Removing the SFT warm-up hurts most.

Each constraint level helps; the semantic and logical constraints contribute most, especially on BIRD. Replacing all constraints with SFT on the same data loses 8.8 and 10.3 points. Single-stage mixing loses 3.6 and 5.3 points, as its reference model lacks the warm-up.

\begin{takeaway}Contrasting references with near-miss errors gives a signal that more SFT data cannot replace, given a warmed-up reference.\end{takeaway}

\subsection{Analysis and Cost}
\label{sec:analysis}\label{sec:cost}
In Figure~\ref{fig:analysis}b, the SFT warm-up mainly helps hard and extra-hard questions (+15.0 and +21.7 points), where the base model often fails to produce well-formed SQL. Constraint training brings the largest gains at each level, especially for medium to extra-hard questions (+31.6 to +43.0 points). As a result, the gap between easy and extra-hard questions shrinks from 53.7 to 9.6 points. Context augmentation adds a further 0.7--2.1 points. Finally, EX varies by at most 1.4 points over the tested values of $\beta$ and $\lambda_{\mathrm{log}}$ (Figure~\ref{fig:sens}).

\method moves the agent cost to training (Figure~\ref{fig:analysis}c,d). Training takes 10.8 GPU hours (2.3$\times$ SFT) and 21.3\,GB of memory, plus 38K GPT-4 calls. Inference is as fast as SFT (0.42\,s per query) and 23$\times$ faster than CogSQL with GPT-4.

\begin{takeaway}Decision-level supervision helps most on multi-clause queries, and building it once with LLM agents is cheaper than calling agents for every query.\end{takeaway}

\section{Conclusion and Future Work}
We propose \method, which enforces the syntactic validity, semantic alignment, and logical consistency of SQL with oracle-verified set-contrastive supervision, and release the \dataset dataset. With 1.7B parameters, \method outperforms larger fine-tuned, RL, and agent-based systems in a single pass.

In future work, we will fine-tune larger open-weight LLMs with \method to study how constraint-based supervision scales with model size, and construct more constraints, e.g., for nested queries and other SQL dialects.

\bibliographystyle{ACM-Reference-Format}
\bibliography{reference}

\appendix
\raggedbottom
\section{Properties of the Set-Contrastive Objective}
\label{app:props}
This appendix explains why Eq.~\ref{eq:unified} is a suitable training objective for all three levels. We show two properties. First, Eq.~\ref{eq:unified} unifies existing losses: the syntactic loss of Eq.~\ref{eq:syntax} and the DPO loss are its special cases (Proposition~\ref{prop:special}), so one objective serves all levels. Second, its gradient pushes down only oracle-certified negatives and fades once a constraint is satisfied (Proposition~\ref{prop:grad}), so equivalent queries are never penalized and training focuses on unsatisfied constraints.

\noindent\textbf{Notation.} We analyze a single constraint $c=(\mathcal{P}_c,\mathcal{N}_c,M_c)$, write $r_y=r_\theta(y,M_c(y))$, and define
\begin{gather*}
    \pi(y)=\frac{e^{\beta r_y}}{\sum_{y'\in\mathcal{P}_c\cup\mathcal{N}_c}e^{\beta r_{y'}}},\qquad \pi(\mathcal{N}_c)=\sum_{y\in\mathcal{N}_c}\pi(y),\\
    \pi_{\mathcal{P}}(y)=\frac{e^{\beta r_y}}{\sum_{y'\in\mathcal{P}_c}e^{\beta r_{y'}}}\quad(y\in\mathcal{P}_c).
\end{gather*}
Intuitively, $\pi(y)$ is the share of candidate $y$ among all candidates, $\pi_{\mathcal{P}}(y)$ is its share among the positives, and $\pi(\mathcal{N}_c)$ is the total share of the negatives, so $\ell(c)$ is small when $\pi(\mathcal{N}_c)$ is small.

The first proposition recovers existing losses from Eq.~\ref{eq:unified}.

\begin{proposition}[Special cases]
\label{prop:special}
(a) At the syntactic level, with candidates $y^{*}_{<t}\oplus v$ for $v\in\mathcal{V}$, $\beta=1$, and $p_{\mathrm{ref}}\equiv1$, Eq.~\ref{eq:unified} equals the syntactic loss of Eq.~\ref{eq:syntax} at position $t$. (b) If $\mathcal{P}_c=\{y^{+}\}$ and $\mathcal{N}_c=\{y^{-}\}$, Eq.~\ref{eq:unified} equals $-\log\sigma\big(\beta(r_{y^{+}}-r_{y^{-}})\big)$, a DPO loss~\cite{rafailov2023dpo} whose reward is restricted to the positions $M_c$.
\end{proposition}
\begin{proof}
(a) With $\beta=1$ and $p_{\mathrm{ref}}\equiv1$, $\exp(r_y)=p_\theta(v\mid x,y^{*}_{<t})$. Since $\mathcal{P}_c\cup\mathcal{N}_c$ covers $\mathcal{V}$, the denominator equals $\sum_{v\in\mathcal{V}}p_\theta(v\mid x,y^{*}_{<t})=1$, and the numerator is the probability mass of $\mathcal{V}_{\mathrm{valid}}(s_t)$. (b) Dividing the numerator and the denominator by $\exp(\beta r_{y^{+}})$ gives $\ell(c)=\log\big(1+\exp(-\beta(r_{y^{+}}-r_{y^{-}}))\big)=-\log\sigma\big(\beta(r_{y^{+}}-r_{y^{-}})\big)$.
\end{proof}

The second proposition characterizes how Eq.~\ref{eq:unified} updates the masked reward $r_y$ of each candidate.

\begin{proposition}[Gradient]
\label{prop:grad}
$\ell(c)=-\log\big(1-\pi(\mathcal{N}_c)\big)\ge0$, and
\begin{equation}
    \frac{\partial\ell(c)}{\partial r_y}=
    \begin{cases}
        \beta\,\pi(y)\ge0, & y\in\mathcal{N}_c,\\
        \beta\big(\pi(y)-\pi_{\mathcal{P}}(y)\big)\le0, & y\in\mathcal{P}_c.
    \end{cases}
\end{equation}
The total gradient magnitudes over $\mathcal{N}_c$ and $\mathcal{P}_c$ both equal $\beta\,\pi(\mathcal{N}_c)$.
\end{proposition}
\begin{proof}
Write
\begin{equation*}
    \ell(c)=-\log\sum_{y\in\mathcal{P}_c}e^{\beta r_y}+\log\sum_{y\in\mathcal{P}_c\cup\mathcal{N}_c}e^{\beta r_y}.
\end{equation*}
The first claim follows from the definition of $\pi$. Differentiating the second term gives $\beta\pi(y)$ for each candidate, and differentiating the first gives $-\beta\pi_{\mathcal{P}}(y)$ for $y\in\mathcal{P}_c$. Since $\pi(y)=\pi_{\mathcal{P}}(y)\,(1-\pi(\mathcal{N}_c))\le\pi_{\mathcal{P}}(y)$, the sign for $y\in\mathcal{P}_c$ follows. Summing gives $\beta\pi(\mathcal{N}_c)$ over $\mathcal{N}_c$ and $\beta\big(1-(1-\pi(\mathcal{N}_c))\big)=\beta\pi(\mathcal{N}_c)$ over $\mathcal{P}_c$.
\end{proof}

\noindent\textbf{Implications.} Proposition~\ref{prop:grad} has three consequences, which justify the claims in Section~\ref{sec:unified}.
\begin{itemize}
    \item \textbf{Equivalent outputs are not penalized.} Only members of $\mathcal{N}_c$, which an oracle certifies as invalid or result-changing, are pushed down, while every positive is pushed up. An equivalent rewriting of $y^{*}$ is therefore never penalized, unlike in token-level SFT.
    \item \textbf{Bounded, self-annealing updates.} The gradient magnitude is at most $\beta$ and vanishes as $\pi(\mathcal{N}_c)\to0$, so satisfied constraints stop contributing.
    \item \textbf{Localized credit.} Since $r_\theta(y,M_c(y))$ depends only on $M_c(y)$, other positions are supervised by $\mathcal{L}_{\mathrm{SFT}}$ alone.
\end{itemize}

\section{Additional Method Details}
\label{app:method}
\begin{figure}[t]
    \centering
    \includegraphics[width=\columnwidth, trim=13 60 8 55, clip]{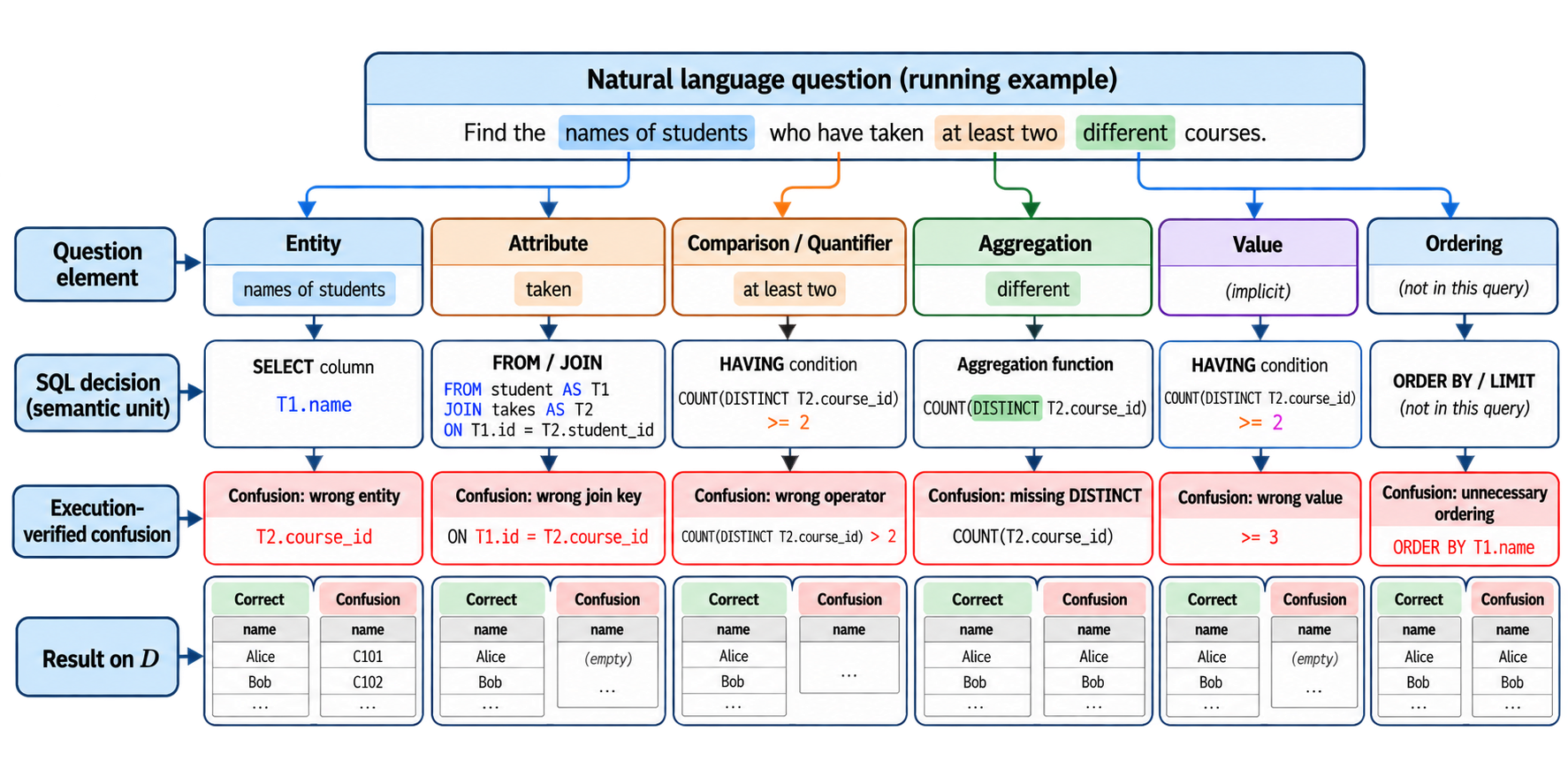}
    \caption{Semantic units of the running example. Each question element is mapped to its SQL decision and contrasted with a confusion (red) that changes the result on $\mathcal{D}$ (bottom row) and is pushed down by the semantic constraint.}
    \Description{Four columns for target entity, quantifier, distinctness, and join, each showing a question phrase, its SQL decision, a confusion, and the results of the correct and confused queries.}
    \label{fig:align}
\end{figure}

\noindent\textbf{Semantic units.} Figure~\ref{fig:align} shows the semantic units of the running example. The six element types are entity (table or join path), attribute (column), value (literal and its column), aggregation (aggregate function and \texttt{DISTINCT}), comparison/quantifier (operator and boundary), and ordering (\texttt{ORDER BY} direction and \texttt{LIMIT}).

\noindent\textbf{Dependency edges.} Table~\ref{tab:dep} lists the edges of the component dependency graph $G$. They are derived from SQL semantics and the key constraints of the schema.

\begin{table}[t]
  \centering
  \tabstyle
  \caption{Dependency edges of $G$, the components each edit affects, and the check the propagated query must pass.}
  \label{tab:dep}
  \begin{tabular}{@{}L{2.45cm}L{2.3cm}L{2.85cm}@{}}
    \toprule
    \headrow \textbf{Edit $\delta$ on} & \textbf{Dependent} & \textbf{Required check} \\
    \midrule
    join path (1:N added) & aggregation, \texttt{SELECT} & duplicate counting; \texttt{DISTINCT} or key \\
    target entity & \texttt{GROUP BY}, \texttt{SELECT}, join & grouping granularity \\
    quantifier & \texttt{HAVING}, sub-query & ``at least'' / ``all'' / ``none'' \\
    boundary / comparison & \texttt{WHERE}, \texttt{HAVING} & inclusive vs.\ exclusive \\
    superlative / top-$k$ & \texttt{ORDER BY}, \texttt{LIMIT}, aggregation & ties; \texttt{MAX} vs.\ sort \\
    filter scope & \texttt{WHERE} vs.\ \texttt{HAVING} & pre- vs.\ post-aggregation \\
    \bottomrule
  \end{tabular}
\end{table}

\section{Implementation Details}
\label{app:impl}
\textbf{Training.} Table~\ref{para:1} lists the hyperparameters and Algorithm~\ref{alg:train} (Section~\ref{sec:schedule}) summarizes the training procedure. LoRA adapters are applied to all attention and feed-forward projections.

\begin{table}[t]
  \centering
	\tabstyle
	\caption{Hyperparameters of \method, selected on a held-out 10\% of the training set.}
	\label{para:1}
	\begin{tabular}{@{}lL{4.1cm}@{}}
		\toprule
		\headrow \textbf{Name} & \textbf{Value} \\ \midrule
		Base LLM & Qwen3-1.7B \\
		Agent LLM & GPT-4 \\
		Fine-tuning method & LoRA, rank 16, $\alpha$ 32, dropout 0.05 \\
		LoRA target modules & q, k, v, o, gate, up, down projections \\
		Learning rate / batch size & 5e-5 / 1 \\
		Epochs (Stage 1 / 2 / 3) & 1 / 2 / 2 \\
		Correction iterations $i^{c}_{\max}$ & 5 \\
		Replay ratio $p_{\mathrm{sample}}$ & 0.1 \\
		Interpolation rate $\gamma$ & 0.5 \\
		Constraint weights $\lambda_{\mathrm{syn}},\lambda_{\mathrm{sem}},\lambda_{\mathrm{log}}$ & 0.5, 1.0, 1.0 \\
		Temperature $\beta$ & 0.1 \\
		Counterexample search per pair & 10 sub-databases $\times$ 200 mutations \\
		Seeds & 3 \\
		Retrieved examples $k$ & 3 \\
		Retrieval embedding / store & nomic-embed-text / Chroma \\
		\bottomrule
	\end{tabular}
\end{table}

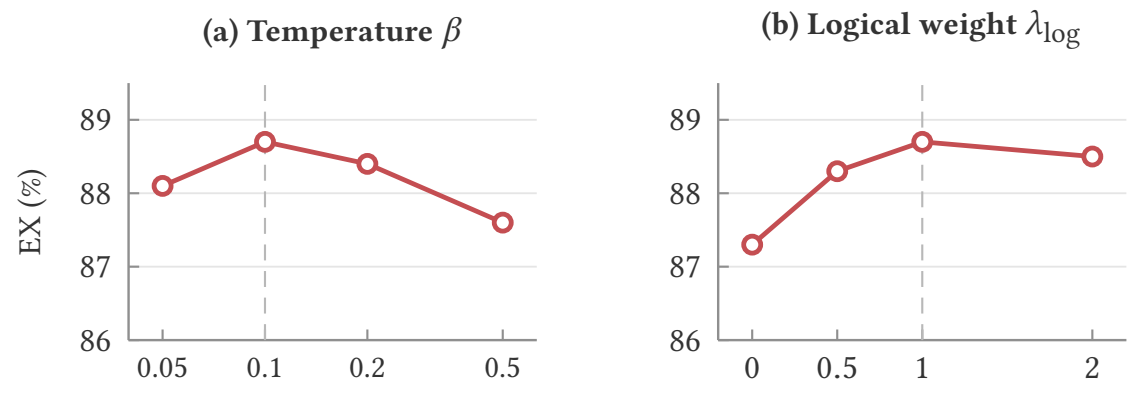
\begin{figure}[t]
    \centering
    \begin{tikzpicture}
\begin{groupplot}[
    group style={group size=2 by 1, horizontal sep=12mm},
    paperstyle,
    scale only axis, width=2.7cm, height=1.7cm,
    ymin=86, ymax=89.5, ytick={86,87,88,89}, ymajorgrids,
    every axis plot/.append style={line width=0.9pt, mark size=1.6pt},
]
\nextgroupplot[title={(a) Temperature $\beta$}, ylabel={EX (\%)}, xmode=log, xtick={0.05,0.1,0.2,0.5}, xticklabels={0.05,0.1,0.2,0.5}]
\draw[cInk!35, densely dashed, line width=0.4pt] (axis cs:0.1,86) -- (axis cs:0.1,89.5);
\addplot[cB, mark=*, mark options={fill=white}] coordinates {(0.05,88.1) (0.1,88.7) (0.2,88.4) (0.5,87.6)};
\nextgroupplot[title={(b) Logical weight $\lambda_{\mathrm{log}}$}, xtick={0,0.5,1,2}]
\draw[cInk!35, densely dashed, line width=0.4pt] (axis cs:1,86) -- (axis cs:1,89.5);
\addplot[cB, mark=*, mark options={fill=white}] coordinates {(0,87.3) (0.5,88.3) (1,88.7) (2,88.5)};
\end{groupplot}
\end{tikzpicture}
    \caption{Sensitivity of Spider Dev EX to the temperature $\beta$ and the logical weight $\lambda_{\mathrm{log}}$.}
    \Description{Two line charts of execution accuracy against beta and lambda_log.}
    \label{fig:sens}
\end{figure}

\noindent\textbf{Hyperparameter selection.} We hold out 10\% of the training set and grid-search $\lambda_{\mathrm{syn}},\lambda_{\mathrm{sem}},\lambda_{\mathrm{log}}\in\{0.1,0.5,1.0,2.0\}$ and $\beta\in\{0.05,0.1,0.2,0.5\}$ with the other values fixed. Figure~\ref{fig:sens} shows the sensitivity to $\beta$ and $\lambda_{\mathrm{log}}$ on Spider Dev.

\begin{table*}[t]
  \centering  \tabstyle
  \caption{Counterfactual pairs per edge type. $q'$ is the edited question, $y'$ the propagated query, and $\tilde{y}'$ the partially propagated query that updates only the edited clause.}
  \label{tab:depex}
  \setlength{\tabcolsep}{2pt}\begin{tabular}{@{}L{1.3cm}L{4.4cm}L{6.6cm}L{5.0cm}@{}}
    \toprule
    \headrow \textbf{Edge} & \textbf{Edited question $q'$} & \textbf{Propagated $y'$} & \textbf{Partially propagated $\tilde{y}'$} \\
    \midrule
    join path & Courses with a lab section per student? & \texttt{\ldots JOIN section \ldots COUNT(DISTINCT T.course\_id)} & same join, keeps \texttt{COUNT(*)} \\
    target entity & Students per \emph{department}? & \texttt{\ldots COUNT(DISTINCT T.id) \ldots GROUP BY D.dept\_id} & keeps \texttt{GROUP BY T.id} \\
    quantifier & Students who passed \emph{all} courses. & \texttt{\ldots NOT EXISTS (\ldots T.grade = 'F')} & keeps \texttt{WHERE T.grade <> 'F'} (any passed) \\
    boundary & Courses with \emph{at least} 4 credits. & \texttt{WHERE credits >= 4} & keeps \texttt{WHERE credits > 4} \\
    top-$k$ & Top student in \emph{each} course. & \texttt{\ldots grade = (SELECT MAX(grade) \ldots)} & keeps \texttt{ORDER BY grade DESC LIMIT 1} \\
    filter scope & Departments with over 100 enrollments \emph{in total}. & \texttt{\ldots GROUP BY dept\_id HAVING COUNT(*) > 100} & keeps the condition in \texttt{WHERE} \\
    \bottomrule
  \end{tabular}
\end{table*}

\noindent\textbf{Oracle implementation.} The grammar oracle is an incremental parser for the SQLite dialect, implemented from scratch. It keeps one scope table per nested query, which records the tables, aliases, and visible columns, and covers nested queries, set operations, and CTEs. Queries outside this grammar are discarded. The execution oracle runs queries in SQLite 3.42.0 with a timeout of 5 seconds. Two queries return the same result if their outputs are equal as multisets of rows, where row order matters only when the reference contains \texttt{ORDER BY}, floating-point values are compared up to 2 decimal places, and \texttt{NULL} values are equal to each other. A query that fails or times out is treated as invalid, and its candidate is discarded. For each query pair, the counterexample generator tries up to 10 sub-databases with at most 200 mutations each. A different result on a database that satisfies the key and type constraints proves that two queries are not equivalent, so the negatives of the semantic and logical constraints are wrong by construction. Positives rest on other evidence. Semantic positives are the references, and a logical positive $y'$ must execute on all databases and pass verification.

\begin{figure}[t]
    \centering
    \includegraphics[width=\columnwidth, trim=22 68 19 27, clip]{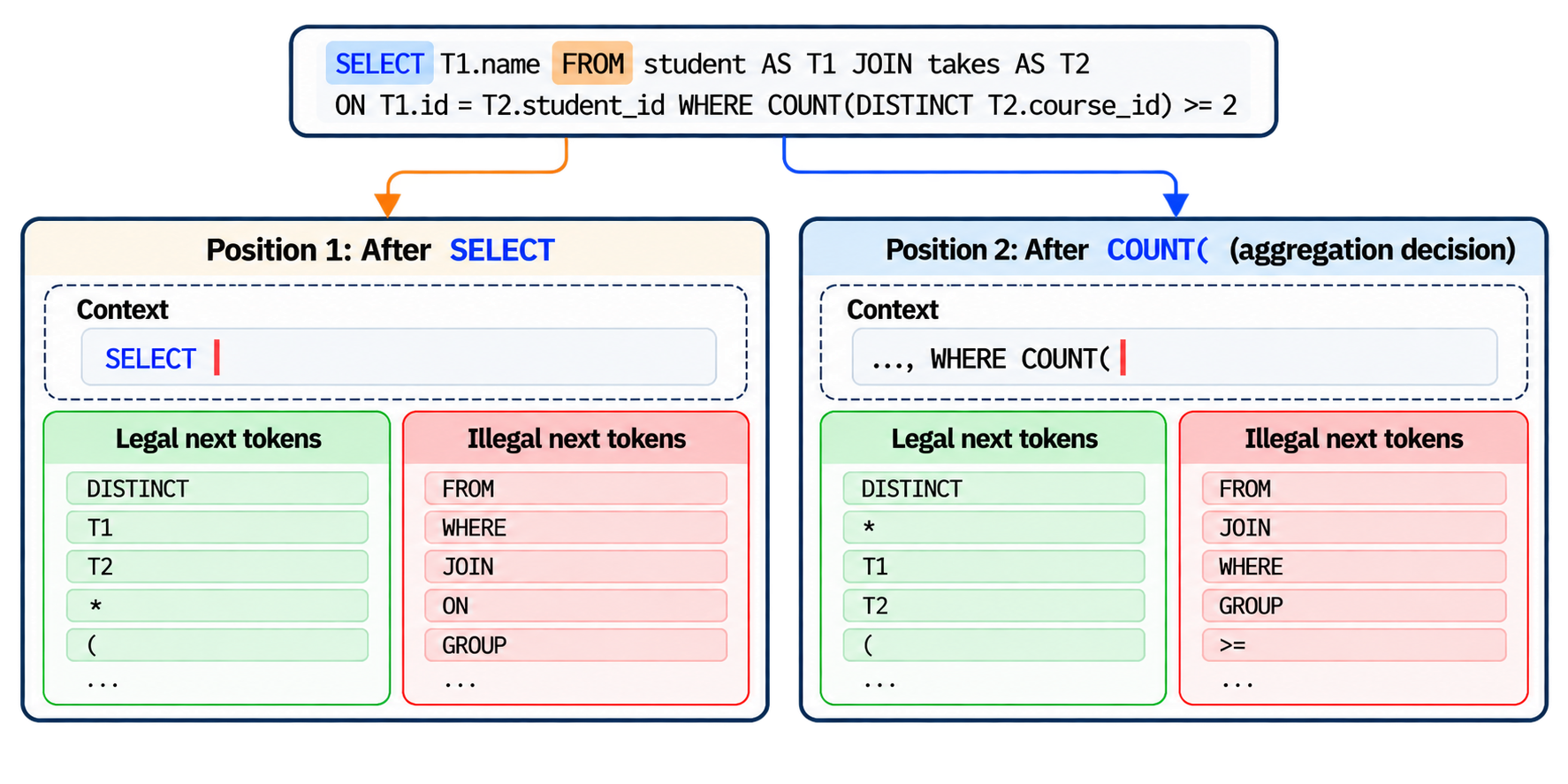}
    \caption{Legal and illegal next tokens derived by the grammar oracle at two decision positions of the running example.}
    \Description{A SQL query split into tokens, with boxes listing valid and invalid tokens at two positions.}
    \label{fig:parse}
\end{figure}

\noindent\textbf{Valid-token sets.} Figure~\ref{fig:parse} shows the legal and illegal tokens that the grammar oracle derives at two decision positions.

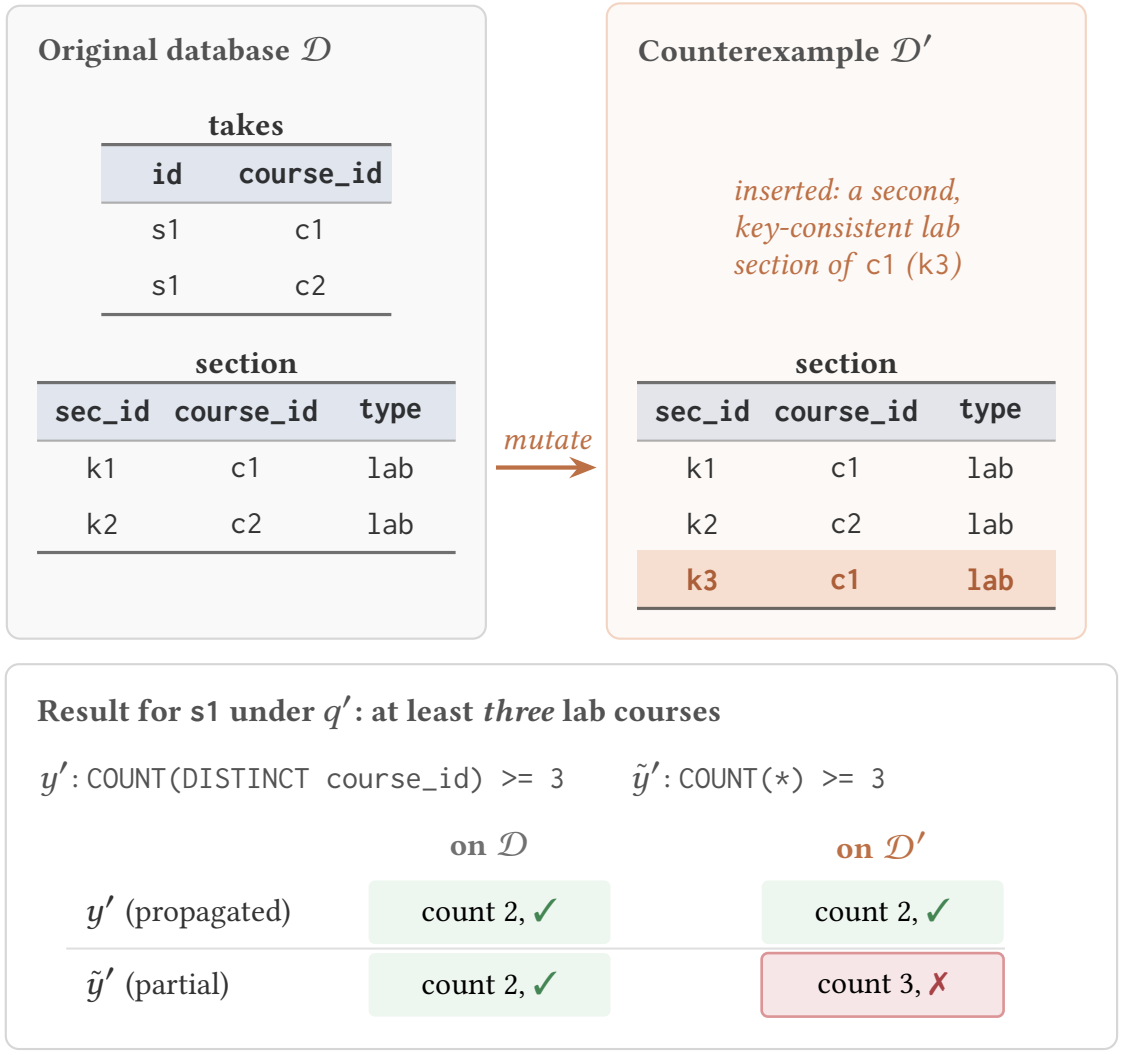
\begin{figure}[t]
    \centering
\begin{tikzpicture}[
    font=\scriptsize,
    tbl/.style={matrix of nodes, column sep=0pt, row sep=0pt, inner sep=0pt,
                nodes={font=\ttfamily\scriptsize, inner sep=1.5pt, minimum width=8.5mm, minimum height=3.7mm, anchor=center, text=cInk},
                row 1/.style={nodes={font=\ttfamily\scriptsize\bfseries}}},
    ttl/.style={font=\scriptsize\bfseries, text=cInk, anchor=south, inner sep=1.5pt},
    panel/.style={draw=cInk!20, rounded corners=2.5pt, line width=0.4pt, inner sep=2mm},
    hdr/.style={font=\scriptsize\bfseries, text=cInk!85, anchor=south west, inner sep=0pt},
    ar/.style={-{Stealth[length=1.8mm, width=1.4mm]}, line width=0.8pt, lvLog!85!black},
    res/.style={font=\scriptsize, minimum width=16mm, minimum height=4.2mm, inner sep=1pt, rounded corners=1.2pt},
  ]
  \newcommand{\ok}{\textcolor{cC!80!black}{\ding{51}}}
  \newcommand{\bad}{\textcolor{cB!85!black}{\ding{55}}}
  \newcommand{\booktab}[2]{%
    \fill[cA!14, blend mode=multiply] (#1-1-1.north west) rectangle (#1-1-#2.south east);
    \draw[cInk!70, line width=0.6pt] (#1.north west) -- (#1.north east);
    \draw[cInk!40, line width=0.4pt] (#1-1-1.south west) -- (#1-1-#2.south east);
    \draw[cInk!70, line width=0.6pt] (#1.south west) -- (#1.south east);}

  \matrix[tbl] (tk) at (0,0) {
    id & course\_id \\
    s1 & c1 \\
    s1 & c2 \\
  };
  \booktab{tk}{2}
  \node[ttl] at (tk.north) {takes};
  \matrix[tbl, anchor=north] (sec) at ($(tk.south)+(0,-4.5mm)$) {
    sec\_id & course\_id & type \\
    k1 & c1 & lab \\
    k2 & c2 & lab \\
  };
  \booktab{sec}{3}
  \node[ttl] at (sec.north) {section};
  \node[hdr] (hD) at ($(sec.west |- tk.north)+(0,5mm)$) {Original database $\mathcal{D}$};

  \matrix[tbl, anchor=north west] (sec2) at ($(sec.north east)+(12mm,0)$) {
    sec\_id & course\_id & type \\
    k1 & c1 & lab \\
    k2 & c2 & lab \\
    |[font=\ttfamily\scriptsize\bfseries, text=lvLog!80!black]| k3 & |[font=\ttfamily\scriptsize\bfseries, text=lvLog!80!black]| c1 & |[font=\ttfamily\scriptsize\bfseries, text=lvLog!80!black]| lab \\
  };
  \booktab{sec2}{3}
  \fill[lvLog!22, blend mode=multiply] (sec2-4-1.north west) rectangle (sec2-4-3.south east);
  \node[ttl] at (sec2.north) {section};
  \node[hdr] (hD2) at ($(sec2.west |- hD.south)$) {Counterexample $\mathcal{D}'$};
  \node[text=lvLog!85!black, align=center, font=\scriptsize\itshape] (note) at (sec2.north |- tk.center) {inserted: a second,\\key-consistent lab\\section of \texttt{\upshape c1} (\texttt{\upshape k3})};

  \begin{scope}[on background layer]
    \node[panel, fill=cInk!3, fit=(hD)(tk)(sec)(sec2.south -| sec.west)] (pD) {};
    \node[panel, fill=lvLog!5, draw=lvLog!35, fit=(hD2)(sec2)(note)] (pD2) {};
  \end{scope}
  \draw[ar] ($(pD.east |- sec.center)+(0.6mm,0)$) -- node[above, text=lvLog!85!black, font=\scriptsize\itshape] {mutate} ($(pD2.west |- sec.center)+(-0.6mm,0)$);

  \node[hdr] (hR) at ($(pD.south west)+(2mm,-6mm)$) {Result for \texttt{s1} under $q'$: at least \emph{three} lab courses};
  \node[anchor=north west, inner sep=0pt, text=cInk!85] (defs) at ($(hR.south west)+(0,-2mm)$) {$y'$: \texttt{COUNT(DISTINCT course\_id) >= 3}\qquad $\tilde{y}'$: \texttt{COUNT(*) >= 3}};
  \coordinate (c1) at ($(hR.west)+(30mm,0)$);
  \coordinate (c2) at ($(hR.west)+(56mm,0)$);
  \node[font=\scriptsize\bfseries, text=cInk!70, anchor=north] (h1) at ($(c1 |- defs.south)+(0,-1.2mm)$) {on $\mathcal{D}$};
  \node[font=\scriptsize\bfseries, text=lvLog!85!black, anchor=north] (h2) at ($(c2 |- defs.south)+(0,-1.2mm)$) {on $\mathcal{D}'$};
  \node[anchor=west, text=cInk] (q1) at ($(hR.west |- h1.south)+(2mm,-2.6mm)$) {$y'$ (propagated)};
  \node[anchor=west, text=cInk] (q2) at ($(q1.west)+(0,-4.8mm)$) {$\tilde{y}'$ (partial)};
  \node[res, fill=cC!10] at (c1 |- q1) {count 2, \ok};
  \node[res, fill=cC!10] at (c2 |- q1) {count 2, \ok};
  \node[res, fill=cC!10] at (c1 |- q2) {count 2, \ok};
  \node[res, fill=cB!14, draw=cB!60, line width=0.5pt] (bad) at (c2 |- q2) {count 3, \bad};
  \draw[cInk!15, line width=0.4pt] ($(q1.west)+(0,-2.4mm)$) -- ($(q1.west -| bad.east)+(0,-2.4mm)$);
  \begin{scope}[on background layer]
    \node[panel, fill=white, fit=(hR)(defs)(q2)(bad)(h2)(pD2.east |- hR)] {};
  \end{scope}
\end{tikzpicture}
    \caption{A counterexample database for the join-path edit. The inserted lab section (highlighted) makes $\tilde{y}'$ wrongly return \texttt{s1} (\ding{55}), while $y'$ stays correct (\ding{51}).}
    \Description{Two small tables, takes and section, with one inserted row highlighted, and the results of two queries.}
    \label{fig:cex}
\end{figure}

\noindent\textbf{Counterexample databases.} Figure~\ref{fig:cex} shows a counterexample database for the join-path edit. The original data cannot distinguish the propagated and partially propagated queries; inserting a second, key-consistent lab section of the same course separates them.

\noindent\textbf{General-purpose LLMs.} All general-purpose LLMs in Section~\ref{sec:llms} receive the same prompt, consisting of the schema in DDL form with sample rows, the question, and an instruction to output a single SQLite query. The first SQL code block of the output is evaluated.

\section{Additional Results}
\label{app:more}\label{app:errors}
\begin{figure}[t]
    \centering
\begin{tikzpicture}
\begin{axis}[
    paperstyle,
    xbar, bar width=5pt,
    scale only axis, width=4.6cm, height=2.35cm,
    xmin=0, xmax=185, xtick={0,50,100,150},
    xlabel={\# cases},
    symbolic y coords={Correct but flagged, Extra columns, Missing columns, Logical error, Syntax error, Wrong column, Semantic mismatch, Over-generic query, Spurious ordering},
    ytick=data,
    y dir=reverse, enlarge y limits=0.06,
    xmajorgrids, y axis line style={draw=none}, ytick style={draw=none},
    nodes near coords, every node near coord/.append style={anchor=west, xshift=-1pt},
]
\addplot[fill=cA!75, draw=none] coordinates {(163,Correct but flagged) (71,Extra columns) (37,Missing columns) (31,Logical error) (18,Syntax error) (17,Wrong column) (15,Semantic mismatch) (14,Over-generic query) (12,Spurious ordering)};
\end{axis}
\end{tikzpicture}
    \caption{Annotation issues in Spider flagged by the error-mining agent and confirmed manually.}
    \Description{Horizontal bar chart of the number of cases for nine issue types.}
    \label{fig:rv}
\end{figure}
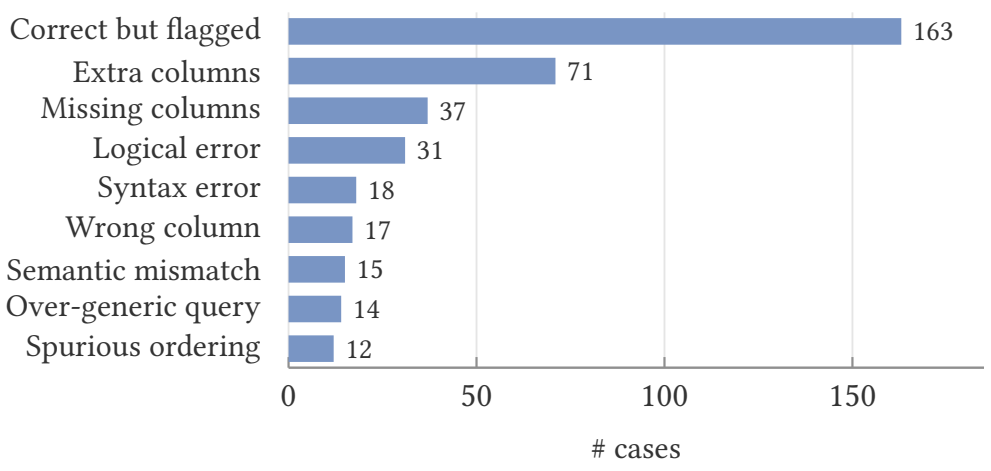

Figure~\ref{fig:rv} lists the Spider annotation issues confirmed while building \dataset (Section~\ref{sec:data}). Figure~\ref{fig:sizeacc} plots the EX of fine-tuning and preference or RL methods against model size (Section~\ref{sec:main}). Table~\ref{tab:errcat} defines the error categories of Section~\ref{sec:motivation}.

\begin{figure}[t]
    \centering
\begin{tikzpicture}[lab/.style={font=\scriptsize, text=cInk!85, inner sep=1pt}]
\begin{groupplot}[
    group style={group size=2 by 1, horizontal sep=9mm},
    paperstyle,
    scale only axis, width=3.05cm, height=1.85cm,
    xmode=log, xmin=1.2, xmax=110, xtick={2,7,70}, xticklabels={2B,7B,70B},
    xlabel={Parameters (log scale)},
    ymajorgrids, xmajorgrids,
    every axis plot/.append style={only marks},
]
\nextgroupplot[title={Spider Test EX (\%)}, ymin=73, ymax=92, ytick={75,80,85,90}]
\addplot[mark=*, mark size=1.7pt, cA!80, fill=cA!60] coordinates {(3,75.1) (3,79.9) (7,83.5) (7,83.5) (7,83.7) (7,87.9) (7,86.8)};
\addplot[mark=square*, mark size=1.6pt, cC!85!black, fill=cC!70] coordinates {(70,86.6) (7,88.7) (7,88.8)};
\addplot[mark=star, mark size=3.4pt, cB!85!black, line width=0.8pt] coordinates {(1.7,89.2)};
\node[lab, anchor=north west, text=cB!85!black, font=\scriptsize\bfseries] at (axis cs:1.25,88.4) {\method};
\node[lab, anchor=west] at (axis cs:3.2,75.1) {PICARD};
\node[lab, anchor=west] at (axis cs:3.2,79.9) {RESDSQL};
\node[lab, anchor=west] at (axis cs:7.6,89.3) {Arctic, SQL-R1};
\node[lab, anchor=west] at (axis cs:7.6,87.6) {OmniSQL};
\node[lab, anchor=west] at (axis cs:7.6,86.2) {JOLT-SQL};
\node[lab, anchor=west] at (axis cs:7.6,83.4) {CodeS, ROUTE, SENSE};
\node[lab, anchor=east] at (axis cs:66,86.6) {ExCoT};
\nextgroupplot[title={BIRD Dev EX (\%)}, ymin=48, ymax=77, ytick={50,60,70}]
\addplot[mark=*, mark size=1.7pt, cA!80, fill=cA!60] coordinates {(7,57.2) (7,51.8) (7,55.9) (7,63.9) (7,60.4)};
\addplot[mark=square*, mark size=1.6pt, cC!85!black, fill=cC!70] coordinates {(70,68.5) (7,66.6) (7,68.9)};
\addplot[mark=star, mark size=3.4pt, cB!85!black, line width=0.8pt] coordinates {(1.7,73.4)};
\node[lab, anchor=north west, text=cB!85!black, font=\scriptsize\bfseries] at (axis cs:1.25,72.0) {\method};
\node[lab, anchor=west] at (axis cs:7.6,69.4) {Arctic};
\node[lab, anchor=west] at (axis cs:7.6,66.4) {SQL-R1};
\node[lab, anchor=west] at (axis cs:7.6,63.9) {OmniSQL};
\node[lab, anchor=west] at (axis cs:7.6,51.8) {SENSE};
\node[lab, anchor=east] at (axis cs:66,68.5) {ExCoT};
\end{groupplot}
\node[anchor=north, font=\scriptsize, text=cInk] at ($(group c1r1.south)!0.5!(group c2r1.south)+(0,-7mm)$) {%
  \tikz[baseline=-0.6ex]\fill[cA!60, draw=cA!80] (0,0) circle (1.7pt);~Fine-tuning\quad
  \tikz[baseline=-0.6ex]\fill[cC!70, draw=cC!85!black] (-1.6pt,-1.6pt) rectangle (1.6pt,1.6pt);~Preference / RL\quad
  \tikz[baseline=-0.6ex]\node[star, star points=5, fill=cB!85!black, inner sep=0pt, minimum size=6pt]{};~\method};
\end{tikzpicture}
    \caption{EX versus model size. Agent-based methods are omitted because the size of GPT-4 is undisclosed.}
    \Description{Two scatter plots of execution accuracy against the number of parameters on a log scale.}
    \label{fig:sizeacc}
\end{figure}
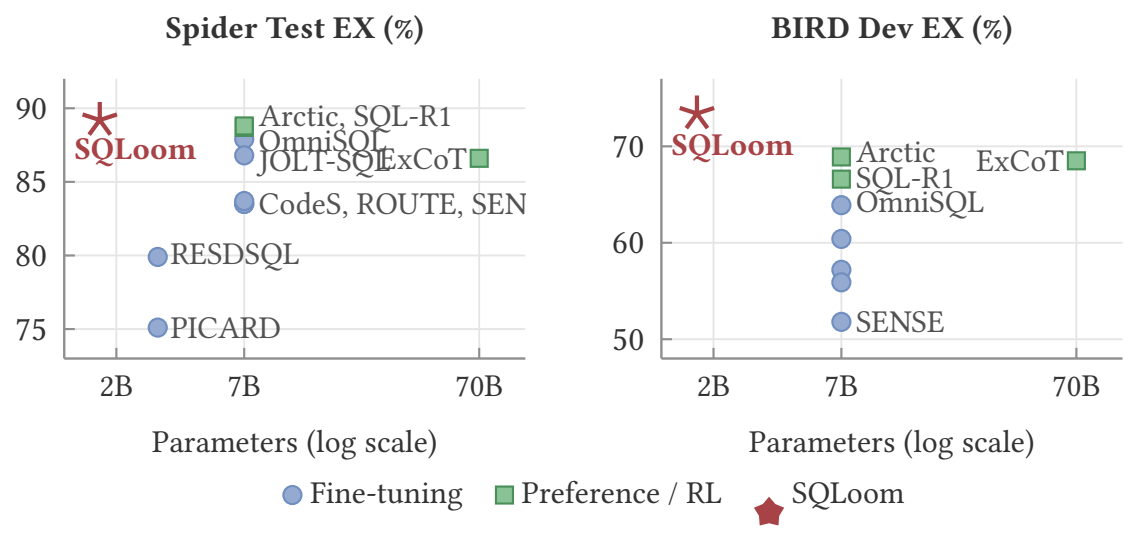

\begin{table}[t]
  \centering
  \tabstyle
  \caption{Error categories used in Figure~\ref{fig:errors}.}
  \label{tab:errcat}
  \setlength{\tabcolsep}{3pt}
  \begin{tabular}{@{}L{1.45cm}L{2.85cm}L{3.6cm}@{}}
    \toprule
    \headrow \textbf{Type} & \textbf{Definition} & \textbf{Example} \\
    \midrule
    Wrong table & wrong or misunderstood table & \texttt{takes} instead of \texttt{student} \\
    Wrong column & mismatched, wrong-key, or non-existent column & \texttt{course\_id} when the title is asked \\
    Semantic error & misread question, wrong alias or aggregate, or missing condition & \texttt{AVG} for ``total'' \\
    Wrong ordering or \texttt{LIMIT} & wrong \texttt{LIMIT} or sort direction & \texttt{ASC} for ``highest'' \\
    Other model error & e.g., hard-coded outputs & returns a constant \\
    \bottomrule
  \end{tabular}
\end{table}

\section{Examples of Auxiliary Supervision}
\label{app:examples}

Table~\ref{tab:confusion} lists the confusion types used by the semantic constraint, and Table~\ref{tab:depex} gives one counterfactual pair for each dependency edge of Table~\ref{tab:dep}. All examples use the \texttt{college} schema with tables \texttt{student(id, name, dept\_id)}, \texttt{takes(id, course\_id, grade)}, \texttt{course(course\_id, title, dept\_id, credits)}, \texttt{section(sec\_id, course\_id, type)}, and \texttt{department(dept\_id, dept\_name)}.

\begin{table}[t]
  \centering  \tabstyle
  \caption{Confusion types of the semantic constraint. Each confusion replaces one decision of the reference query.}
  \label{tab:confusion}
  \begin{tabular}{@{}L{1.6cm}L{3.0cm}L{3.0cm}@{}}
    \toprule
    \headrow \textbf{Element} & \textbf{Reference decision} & \textbf{Confusion} \\
    \midrule
    entity & \texttt{FROM student} & \texttt{FROM takes} \\
    attribute & \texttt{course.title} & \texttt{course.course\_id} \\
    value & \texttt{dept\_name = 'Physics'} & \texttt{dept\_name = 'physics'} \\
    aggregation & \texttt{COUNT(DISTINCT course\_id)} & \texttt{COUNT(*)} \\
    comparison & \texttt{credits >= 4} & \texttt{credits > 4} \\
    quantifier & \texttt{NOT EXISTS (\ldots)} & \texttt{NOT IN} with nullable key \\
    ordering & \texttt{ORDER BY grade DESC LIMIT 1} & \texttt{ORDER BY grade ASC LIMIT 1} \\
    \bottomrule
  \end{tabular}
\end{table}

\section{Prompt Templates}
\label{app:prompts}
Each agent prompt starts with the schema and sample rows of the database as DDL and DML statements, together with the schema descriptions of \dataset, and then lists the inputs below. Prompts~\ref{tab:p_eval}--\ref{tab:p_verify} show the agent templates. The fine-tuned model is asked to translate the question into one SQLite query and output only the SQL. Its input has the schema with descriptions, the question, and, in Stage 3, retrieved pairs from the same database.

\begin{promptbox}[label=tab:p_eval]{Templates of the error-mining agent $A_e$}
\textbf{Evaluation.} You are a database expert. Determine whether the labeled SQL or the predicted SQL better aligns with the given question, based on the provided database schema and content. \textbf{Correction.} Given the question and an erroneous SQL query, correct the SQL while preserving its structure as closely as possible.

\smallskip\textbf{Input.} Question: \texttt{\{question\}}. Labeled SQL: \texttt{\{label\_SQL\}}. Predicted SQL: \texttt{\{predict\_SQL\}}.
\end{promptbox}

\begin{promptbox}[label=tab:p_align]{Template of the alignment agent $A_a$}
\textbf{System.} You are a database expert. Given a question and its SQL query, list each semantic element of the question (entity, attribute, value, aggregation, comparison or quantifier, ordering) and the SQL fragment that realizes it. For each fragment, propose up to three alternatives of the same type that a model may confuse with it. Output one line per element in the form \texttt{element -> fragment | alternatives}.

\smallskip\textbf{Input.} Question: \texttt{\{question\}}. SQL: \texttt{\{label\_SQL\}}. Frequent errors of the current model: \texttt{\{error\_decisions\}}.
\end{promptbox}
\begin{promptbox}[label=tab:p_edit]{Template of the editing agent $A_d$}
\textbf{System.} You are a database expert. Given a question, its SQL query, and an edit type, rewrite the question so that it expresses the edited intent. Output (1) the edited question, (2) the correct SQL in which all clauses affected by the edit are updated, and (3) a SQL that applies the edit only to the targeted clause and copies all other clauses from the original SQL.

\smallskip\textbf{Input.} Question: \texttt{\{question\}}. SQL: \texttt{\{label\_SQL\}}. Edit type and dependent clauses: \texttt{\{edge\}}.
\end{promptbox}
\begin{promptbox}[label=tab:p_verify]{Template of the verification agent $A_v$}
\textbf{System.} You are a database expert. Given a SQL query, write one natural language question that the query answers. Then decide whether your question has the same meaning as the given question. Answer \texttt{yes} or \texttt{no} with a short reason.

\smallskip\textbf{Input.} SQL: \texttt{\{edited\_SQL\}}. Question: \texttt{\{edited\_question\}}.
\end{promptbox}

\end{document}